\documentclass[AMS,Times1COL]{WileyNJDv5}

\usepackage{multicol}
\usepackage{multirow}
\usepackage{caption}
\usepackage{listings}
\usepackage{xcolor}
\usepackage{enumitem}
\usepackage{algpseudocode}

\algnewcommand{\LineComment}[1]{\Statex \(\triangleright\) #1}

\definecolor{olivegreen}{HTML}{267F99}
\newcommand{\silence}[1]{}

\newcommand{\valpct}[2]{#1 (#2)}

\usepackage{makecell}

\lstdefinelanguage{JavaScript}{
  basicstyle=\ttfamily\scriptsize,
  keywords={typeof, new, true, false, catch, function, return, null, catch, switch, var, if, in, while, do, else, case, break, import, export, {*}, as, from, this},
  keywordstyle=\color{blue},
  ndkeywords={class, export, boolean, throw, implements, import, this},
  ndkeywordstyle=\color{darkgray}\bfseries,
  identifierstyle=\color{black},
  sensitive=false,
  comment=[l]{//},
  morecomment=[s]{/*}{*/},
  commentstyle=\color{olivegreen}\ttfamily,
  stringstyle=\color{purple}\ttfamily,
  morestring=[b]',
  morestring=[b]"
}

\articletype{Article Type}%

\startpage{1}

\begin{document}

\title{Augmenting unit test suites from integration tests}

\author[1]{Katerina Paltoglou}
\author[2]{Vassilis E. Zafeiris}

\address[1]{\orgdiv{Department of Informatics}, 
  \orgname{Athens University of Economics and Business},
\orgaddress{\state{Athens}, \country{Greece}}}

\address[2]{\orgdiv{Computer Science and Engineering Department}, 
  \orgname{University of Ioannina},
\orgaddress{\state{Ioannina}, \country{Greece}}}

\corres{Vassilis E. Zafeiris (bzafiris@uoi.gr)}


\abstract[Abstract]{
We propose a method that employs static and dynamic analysis
for augmenting a test suite with automatically generated unit tests.
The method is most suitable for test suites
where the stratification of unit, integration and system tests 
does not conform to the recommended test pyramid structure:
numerous unit tests providing high code coverage and forming the base,
fewer integration tests in the middle that verify component collaboration,
and far fewer system or UI tests at the top 
that exercise acceptance or other scenarios of use.
Instead, integration and system tests represent the majority of test cases,
resulting in coarse-grained tests with limited fault localization and longer execution times.
The method leverages integration tests, 
exercising a component and its dependencies,
to generate unit tests that verify component dependencies in isolation.
We showcase and empirically evaluate the proposed method in the Node.js platform, 
although it can be ported and adapted to other languages and platforms.
The evaluation is based on a research prototype implemented as a Node.js tool
and is conducted in the context of twelve open source JS applications (benchmark projects).
Evaluation results support the effectiveness and practicality of our approach.
}


\maketitle

\renewcommand\thefootnote{\fnsymbol{footnote}}
\setcounter{footnote}{1}


\newcommand{\Global}[1]{\State \textbf{global} #1}
\newcommand{\Returns}[1]{\State \textbf{return} #1}
\newcommand{\ObjRef}{\code{ObjRef}}
\newcommand{\StackC}{$stack_C$}
\newcommand{\StackS}{$stack_S$}
\newcommand{\code}[1]{\texttt{#1}}
\newcommand{\isObject}{\texttt{isObject}}

\newcommand{\goojs}{\texttt{goojs}}
\newcommand{\planckjs}{\texttt{planck.js}}
\newcommand{\vtree}{\texttt{vtree}}
\newcommand{\stringtemplate}{\texttt{string-template}}

\newcommand{\wavdecoder}{\texttt{wav-decoder}}
\newcommand{\underscorestring}{\texttt{underscore.string}}
\newcommand{\tslibmathjs}{\texttt{ts-lib-mathjs}}
\newcommand{\subsetF}{\texttt{\_setsubset()}}
\newcommand{\m}{\texttt{m}}
\newcommand{\values}{\texttt{\_values}}
\newcommand{\SparseMatrix}{\texttt{SparseMatrix}}
\newcommand{\ar}{\texttt{arg1}}
\newcommand{\indexC}{\texttt{Index}}

\newcommand{\handleThunk}{\texttt{handleThunk()}}
\newcommand{\isThunk}{\texttt{isThunk()}}
\newcommand{\renderThunk}{\texttt{renderThunk()}}
\newcommand{\VirtualNode}{\texttt{VirtualNode}}
\newcommand{\VirtualText}{\texttt{VirtualText}}

\newcommand{\VNode}{\texttt{VNode}}
\newcommand{\vnode}{\texttt{vnode}}

\newcommand{\eval}{\texttt{eval}}

\newcommand{\underscoreString}{\texttt{underscore.string}}
\newcommand{\classifyjs}{\texttt{classify.js}}
\newcommand{\classify}{\texttt{classify()}}
\newcommand{\str}{\texttt{str}}

\section{Introduction} \label{sec:intro}

Automated testing plays a key role in modern software development.
It is the cornerstone of agile methods, 
enhancing the safety of refactorings to enable continuous design
and providing a safety net against regressions in fast development cycles~\cite{shore2008}.
Moreover, it enables continuous integration practices,
ensuring software builds remain constantly in a reliable and deployable state~\cite{humble2010}.
In software engineering practice, 
automated testing is, typically, performed at three levels: 
(i) unit testing with focus on verifying software units isolated from their dependencies,
(ii) integration testing that checks the collaboration among software units or subsystems
and, (iii) system or end-to-end testing that exercises the system as a whole.
The contribution of each level to the test suite, in terms of number of tests,
is commonly described in the test pyramid~\cite{Cohn09}.
Unit tests form the base of the pyramid, due to their large number
and emphasis on achieving high code coverage.
Integration tests, verifying component collaboration, are fewer and lay in the middle, 
while far fewer system tests are at the top of the pyramid, 
exercising acceptance or other scenarios of use.

Although unit tests constitute the base of the test pyramid, 
often, the logic of a software unit is exercised indirectly
through integration or system tests and not by dedicated unit tests.
An inverted shape of the test pyramid characterized by a limited number of unit tests
has a negative impact to the maintenance and evolution of a code base.
This impact is exacerbated in dynamic programming languages 
characterized by weak typing and the support of dynamic features,
such as runtime manipulation of an object's structure,
where reliance on runtime validation and extensive automated tests is more prominent.

We propose a method that employs static and dynamic analysis
for augmenting a test suite with automatically generated unit tests.
The method is most suitable for test suites
where the stratification of unit, integration and system tests 
does not conform to the recommended test pyramid structure.
Instead, integration and system tests represent the majority of test cases,
resulting in coarse-grained tests with limited fault localization and longer execution times.
Our method leverages integration tests, that exercise a component and its dependencies,
to generate unit tests verifying component dependencies in isolation.
We showcase and empirically evaluate the proposed method in the Node.js platform, 
although it can be ported and adapted to other languages and platforms.

As part of this work we investigate 
the effectiveness and practicality of the proposed test augmentation method,
through an empirical study that provides answers to the following research questions:

\begin{enumerate}
    \item What is the degree of test augmentation attained by the proposed method?
    \item Does the proposed method enhance the fault detection capability of a test suite?
    \item Does the proposed method generate maintainable unit tests?
\end{enumerate} 

The main findings of the study indicate that
the effectiveness of test generation depends
on the prevalence of integration tests in the benchmark projects
and the number of call site of dependencies within the exercised components.
In any case, the study confirmed the generation of several unit tests per integration test,
underpinning the effectiveness of the approach.
Although generated unit tests cover subpaths 
of the execution paths activated by integration tests,
the fault location capability of the test suite is improved.
Actually, each generated test introduces assertions 
corresponding to intermediate program states, 
enabling, thus, detection of faults whose effects are not
propagated to the result of its respective integration test.
This improvement in fault detection capability 
has been empirically validated in our study through mutation analysis.
The practicality of the approach is, also, enhanced 
due to the form of test fixture setup code,
especially, in cases that it involves initialization of complex object structures.
The generated unit tests employ statements from the target component and the integration test
for test fixture setup, resulting to intent-revealing and intuitive code.

The rest of this paper is organized as follows. 
In Section~\ref{method} we specify the proposed test augmentation method.
Section~\ref{evaluation} presents the details 
and results of the empirical evaluation of the method, 
while Section~\ref{threats-to-validity} describes 
relevant threats to validity. 
Section~\ref{relatedWork} discusses related work 
and state-of-the-art tools and approaches that 
are proposed towards unit test generation. 
Finally, the paper is concluded in 
Section~\ref{conclusion}.


\section{Test augmentation method} \label{method}

We propose a method for augmenting a test suite with unit tests
through leveraging existing integration and system tests.
The method employs static and dynamic analysis
for inferring test fixtures and assertions for the generated tests.
We showcase and evaluate the proposed method in the Node.js platform,
although it can be ported and adapted to other languages and platforms.
The method improves fault localization~\cite{xuan2014test, soremekun2021locating},
since each generated unit test verifies a more narrow scope of functionality.
Moreover, it enhances the fault detection capability of a test suite
with tests that verify program states that were internal for the functionality
exercised in original integration tests.

In this work, a component refers to a function or method
whose behaviour is verified through one or more tests.
A component implementation may invoke other components,
also, declared as functions or methods.
Let \(C\) be a component and \(D=\cup\{d_i\}, i\in\{1,\ldots,n\}\)
be the set of components invoked within its implementation.
The components in $D$ will be referred as dependencies of \(C\).
Moreover, let \(T_C\) be a subset of the project's test suite $T$
that comprises test cases exercising the behaviour of \(C\).
Given that \(D\) is non empty,
the tests in \(T_C\) are regarded as integration tests.
Our assumption is that tests in \(T_C\)
do not employ a mocking mechanism to isolate \(C\)
from its dependencies.

Let $S_{i}=\cup\{s_{ij}\}, j \in \{1\ldots,k_i\}$ 
be the call sites of dependency $d_i$
within the implementation of $C$.
Given a test case $t \in T$ and a dependency $d_i \in D$,
our method is capable of generating one unit test
for each reachable call site of $d_i$ under the execution of $t$.
Let $|S| =\sum_{i=1}^{n} k_i$ 
be the total number of call sites for all dependencies of $C$,
then the maximum number of unit tests 
that can be generated by the proposed method 
would be $|S|\times |T_C|$.

\begin{figure*}[htb]
\begin{minipage}[t]{.45\textwidth}
  \begin{lstlisting}[language=JavaScript,
      basicstyle=\ttfamily\footnotesize,
      captionpos=t,label={lst:example-classes}]
// Rectangle declarations
function Rectangle(p1, p2, p3, p4) {
  this.points = [p1, p2, p3, p4]
}

Rectangle.normalize = function(dx, dy) {
  const len = Math.hypot(dx, dy);
  return { x: dx / len, y: dy / len };
}

// Point declarations
function Point(x, y) {
  this.x = x;
  this.y = y;
}

Point.prototype.distanceFrom = 
  function (point) {
    return Math.hypot(
      this.x - point.x, this.y - point.y);
}

Point.prototype.moveAlong = 
  function (direction, distance) {
    this.x += direction.x * distance;
    this.y += direction.y * distance;
}
  \end{lstlisting}
  \end{minipage}%
  \begin{minipage}[t]{.45\textwidth}
  \begin{lstlisting}[language=JavaScript,
      basicstyle=\ttfamily\footnotesize,
      captionpos=t, label={lst:stretchLongestEdge}]
// Rectangle declarations
Rectangle.prototype.stretchLongestEdge = 
  function(amount) {

    let maxLen = -Infinity;
    let edgeIndex = 0;

    // Find longest edge
    for (let i = 0; i < 4; i++) {
      const a = this.points[i];
      const b = this.points[(i + 1) % 4];
      const len = a.distanceFrom(b);

      if (len > maxLen) {
        maxLen = len;
        edgeIndex = i;
      }
    }

    const pA = this.points[edgeIndex];
    const pB = this.points[(edgeIndex + 1) % 4];

    const dx = pB.x - pA.x;
    const dy = pB.y - pA.y;

    const normal = Rectangle.normalize(-dy, dx);
    // stretch across normal direction
    pA.moveAlong(normal, amount)
    pB.moveAlong(normal, amount)

    return {
      stretchedEdge: [edgeIndex, (edgeIndex + 1) % 4],
      originalLength: maxLen,
      amountMoved: amount
    };
}
  \end{lstlisting}
  \end{minipage}
    \caption{Example code for motivating and showcasing the proposed method.}
    \label{fig:rectangle-code}
\end{figure*}

\begin{figure*}[htb]
  \centering
\begin{minipage}[t]{.45\textwidth}
  \begin{lstlisting}[language=JavaScript,
      basicstyle=\ttfamily\footnotesize,
      captionpos=t, caption={Integration test},
      label={lst:integration-test}]
// Test case exercising Rectangle

describe('Rectangle', function () {

    it('should stretch longest edge', function () {

        let p0 = new Point(0, 0);
        let p1 = new Point(0, 4);
        let p2 = new Point(3, 4);
        let p3 = new Point(3, 0);

        let r = new Rectangle(p0, p1, p2, p3)
        const result = r.stretchLongestEdge(2)
        expect(result.originalLength).toBe(4)
        expect(result.stretchedEdge[0]).toBe(0)
        expect(result.stretchedEdge[1]).toBe(1)

    });

});
  \end{lstlisting}
  \end{minipage}%
  \begin{minipage}[t]{.45\textwidth}
  \begin{lstlisting}[language=JavaScript,
      basicstyle=\ttfamily\footnotesize,
      captionpos=t, caption={Generated Unit test},
      label={lst:unit-test}]
// Generated test cases 
describe('Point', function () {

  it('moveAlong', function () {
    // code from test case
    let p0 = new Point(0, 0);
    let p1 = new Point(0, 4);
    let p2 = new Point(3, 4);
    let p3 = new Point(3, 0);

    let r = new Rectangle(p0, p1, p2, p3);
    var amount = 2;
    // code from target component
    var edgeIndex = 0;
    const pA = r.points[edgeIndex];

    const normal = Rectangle.normalize(-4, 0);
    pA.moveAlong(normal, 2);

    expect(pA.x).toBe(-2);
    expect(pA.y).toBe(0);

  });
});
  \end{lstlisting}
  \end{minipage}
    \caption{Integration test for \texttt{Rectangle.stretchLongestEdge()} component 
    and generated unit test for is \texttt{Point.moveAlong()} dependency.}
    \label{fig:rectangle-tests}
\end{figure*}

Figure~\ref{fig:rectangle-code} illustrates a fictitious example
that demonstrates the motivation and benefits of the proposed method.
We elected to introduce a code snippet from a broadly familiar domain
in order to not distract the reader with domain specific details.
On the other hand, the presented case study has adequate complexity
in order to highlight the merits and challenges of the proposed method.
Specifically, Listing~\ref{lst:example-classes} presents the implementations
of \code{Point} and \code{Rectangle} classes, 
along with simple utility methods.
Listing~\ref{lst:stretchLongestEdge} depicts the implementation
of a more complex method of the \code{Rectangle} class
that stretches the rectangle's longest edge across its vertical direction
at an amount provided as method parameter.
The \code{stretchLongestEdge()} method has the role of the target component,
while methods \code{distanceFrom()} and \code{moveAlong()} 
of the \code{Point} class serve as component dependencies.

The \code{stretchLongestEdge} method is exercised
through the test case illustrated in Listing~\ref{lst:integration-test}.
We treat this test case as an integration test,
since the \code{Rectangle} functionality depends on \code{Point} calls
for its fullfillment.
The proposed method enables the automated generation 
of test cases for each reachable method invocation of \code{Point} class
from the test case in Listing~\ref{lst:integration-test}.
Specifically, it can generate:
(a) 4 test cases for the \code{distanceFrom} method,
due to its invocation in a for loop with 4 iterations 
(lines 9--12, Listing~\ref{lst:stretchLongestEdge}) and
(b) 2 test cases for the \code{moveAlong} method,
one for each invocation in lines 28--29 of the same listing.
Listing~\ref{lst:unit-test} presents a test case generated
for the \code{moveAlong} invocation in line 28.

We can observe that the generated test enhances software fault localization,
since it verifies the \code{moveAlong()} functionality separately
from \code{stretchLongestEdge()} preventing propagation of 
errors manifesting in the two former functions. 
This is achieved in terms of (a) narrowing down the source location 
where faults are uncovered, with respect to faults 
uncovered both in the initial and the generated test, 
(b) uncovering faults in the generated test
which were hidden in the initial test, 
through the assertions verifying the state of the \code{Point} instance.

In the rest of this section, 
we specify the proposed method
as a sequence of processing steps leveraging 
static and dynamic analysis. 
The method receives as input arguments:
(a) the production code directory of a Node.js application,
(b) the test code directory with the implementation of the test suite $T$,
and (c) the name of the component $C$ 
which is the target for test generation, and
(d) the file location of $C$.
The method generates its output in the test directory,
comprising a set of JS source files.
The source files implement test cases 
for all dependencies of the \emph{target} component $C$
that are reachable by the original test suite $T$.

The processing steps involved 
in the proposed test augmentation procedure are summarized below:

\begin{enumerate}

  \item \textbf{Resolution of call sites for target component dependencies}: 
  relevant source files in the production directory
  are statically analyzed in order to identify the declarations
  of the target component $C$ and its dependencies $D$.
  For each dependency $d_i \in D$, 
  AST locations $L_{D,i}$ for all its call sites within the implementation of $C$
  are statically resolved and their union $L_D = \cup \{L_{D,i}\}$
  is returned as output of this step.
  
  \item \textbf{Test case filtering}:
  the test suite $T$ is dynamically analyzed
  in order to identify test cases with an execution path
  that reaches at least one call site in $L_D$.
  For each test case $t_r$ that reaches
  a call site $s_{i,j}$ of dependency $d_i$,
  we keep its AST location $loc_r$.
  The rest of the test cases in $T$ are discarded
  from further processing.
  Let $T_C = \cup \{t_r\}$ be the set of filtered test cases
  and $L_T$ be their AST locations.
  Their source code is combined in a single test file
  that will drive the dynamic analysis in the next step.
  
  \item \textbf{Identification of seed execution paths}:
  the test suite $T_C$ is dynamically analyzed
  in order to trace the execution paths that cover the call sites $S$.
  For each execution of a test case $t_r$
  that reaches call sites $S_r \subseteq S$
  we create an appropriate abstraction of its execution path $p_r$,
  comprising representations of statements and their execution context.
  Each statement maintains information concerning 
  assignments of primitive values to variables,
  as well as object references that they use or mutate.
  These execution paths will be referred to as ``seed'' execution paths, 
  since they seed the test generation process with essential information
  for the production of test fixtures and assertions.

  \item \textbf{Tracing of object flow dependencies in seed execution paths}:
  in this step, we trace the object flow dependencies among statements 
  within each seed execution path.
  An object flow dependency among two statements na, nb, 
  where $n_a$ precedes $n_b$ in an execution path, 
  denotes that $n_b$ uses at least one object whose state is mutated in statement $n_a$.
  Object flow dependencies will enable the extraction of program slices, 
  having as seed statements the call sites of dependencies in D. 
  
  \item \textbf{Test generation}: the final step generates test case implementations
  for each seed execution path $p_i \in P$ and and each call site of target component dependencies contained within $p_i$.
  
\end{enumerate}

\subsection{Resolution of call sites for target component dependencies}
\label{sect:call-site-resolution}

The first step of the test augmentation procedure receives as input 
the name $id_C$ of the target component $C$, its file location,
as well as the production and test code directories of the analyzed Node.js application.
The purpose of this step is to identify 
(a) the AST location $loc_C$ of the target component $C$,
(b) the list of AST locations $L_D$ for all call sites
of target component dependencies within the bounds of $loc_C$ and
(c) the list of AST locations $L_T$ for all test cases included in the project's test suite.
Notice that an AST location record comprises
the path of the JS source file declaring the respective code
and the start/end character positions within the file 
that delimit the source of the AST fragment.

The identification of required AST locations 
is based on static analysis of relevant JS source files.
At first, the source file of $C$ is parsed 
to construct its AST representation $ast_F$.
The AST location $loc_C$ is retrieved 
through traversal of $ast_F$ 
in search of a \code{FunctionDeclaration} or \code{MethodDefinition} node
named after identifier $id_C$.
Let $ast_C$ be the identified AST node, 
which is, further, processed in order to identify call sites of component dependencies.

For brevity reasons and better focus on core concepts of the proposed method,
we assume that call sites of component dependencies have the basic form
of a function/method invocation: \\

\code{[(const | var | let | "") result = ] [base.]dependency([arguments])} \\

Thus, in terms of JS AST types, the call site corresponds to either
(a) a \code{VariableDeclaration} statement, 
defining a variable to the result of a function/method invocation (\code{CallExpression}),
(b) an \code{ExpressionStatement}, comprising the assignment of a \code{CallExpression} 
to a variable (\code{AssignmentExpression}), or
(c) an \code{ExpressionStatement} involving a single \code{CallExpression}.
Let \code{result} represent the optional variable identifier 
and \code{base} the optional object identifier.
Moreover, let \code{dependency} be the name of the invoked function and 
\code{arguments} the list of actual parameters.

The generation of list $L_D$ is based on AST traversal of the $ast_C$ subtree
in search of nodes that match the required form of call sites.
The retrieved nodes are, further, filtered in order to keep call sites
with invocation of functions/methods declared outside the source file of target component $C$
and their source files belong to the production code of the analyzed project.
As concerning list $L_T$, it is populated through static analysis 
of source files contained in the test directory and recovery of the AST of all test cases.
Finally, all test cases in $L_T$ are merged into a single test module 
which will serve as the dynamic analysis entry point in the following steps.

\subsection{Test case filtering}
\label{sect:test-filtering}

The purpose of this step is to exclude 
from further processing, 
test cases that are not relevant for unit test generation.
These test cases do not exercise the target component
and their execution path does not reach any call site
of component dependencies in $D$.
Test filtering is based on dynamic analysis of the project's test suite.
Dynamic analysis involves instrumentation of the project's code
through interleaving the execution of code instructions
with invocations of appropriate callback functions.
These functions, will be, henceforth, 
referred to as as ``hook'' functions.

Code instructions represent elementary operations 
such as variable assignment, variable read/write,
function/method invocation etc. 
A program statement may comprise more than one instructions,
e.g. statement \texttt{let x = foo() + bar()}
includes one variable write instruction,
two name dereferencing instructions (read)
and two function invocation instructions.
Each instruction is characterized by a unique numeric id (iid)
that maps directly to the AST location of the instruction.
The instrumentation framework invokes appropriate hooks
before/after the execution of instructions,
statements and functions.

Test case filtering is based on an algorithm,
whose steps are asynchronously executed
as part of the instrumented test suite's hooks.
The algorithm receives as input 
(a) a list of the AST locations $L_T$ for the project's test cases,
(b) a list of the AST locations $L_D$ for all call sites of 
target component dependencies,
(c) the AST location $loc_C$ of the target component $C$.
The algorithm returns as output a set of relevant test methods $T_C$
whose execution reaches call sites of component dependencies.
The sources of the test methods are concatenated in a single test script
that drives the dynamic analysis of the next step.

Algorithm operation depends on utility functions 
that determine the AST location of an instruction $iid$:

\begin{itemize}[leftmargin=1cm]
    \item \texttt{iidToLocation(iid)}:
    returns the AST location of the instruction that corresponds to the given $iid$,

    \item \texttt{belongsToAst(loc, iid)}: 
    returns true in case that instruction $iid$ is within the range
    of the given AST location $loc$. 
    For instance, \texttt{belongsToAst($l_C$, iid)} returns true 
    if instruction $iid$ is part of the target component declaration $C$.
\end{itemize}

The proposed algorithm tracks function/method invocation instructions
and maintains a stack \StackC{} of their execution contexts,
abstracting the call stack.
The execution context $ctx_n$ of an instruction $n$ refers 
to a specific invocation of the function declaration 
that includes the instruction.
Assume that instruction $n$ is executed
as part of a function $f$.
Moreover, let $iid_m$ be the id 
of an invocation instruction $m$ of function $f$
within the execution context $ctx_m$.
Then, $ctx_n$ is defined as a tuple $(iid_m, f, type, ctx_m)$
which references its parent context $ctx_m$.
The $type$ item states whether the execution context
corresponds to: (i) the test script ($ROOT$), 
(ii) a test function ($TEST$),
(iii) the target component ($CMP$) or one of its dependencies ($DEP$),
(iv) any other component ($OTHER$).
Thus, execution contexts for a given test function $t_r$
form a tree rooted at the execution context $ctx_r$
of the script that calls $t_r$ 
and, potentially, other test cases.

\begin{algorithm}
\caption{\enskip Test filtering algorithm}\label{alg:test-filter}
\scriptsize
\begin{algorithmic}[1]
  \State INPUT: $L_T$, $L_D$, $loc_C$
  \LineComment{Set of AST locations of relevant test cases}
  \State OUTPUT: $T_C$ 
  
  \State $ invId \gets null$

  \LineComment{Parent execution context for current instructions}
  \State $ctx_p \gets (1, null, ROOT, null)$ \Comment{Initialized to root context}
  \State \Call{push}{\StackC{}, $ctx_p$}

  \Statex
  \Function{onInvokeFunPre}{iid, f, base, args, result}
    \LineComment{Keep the iid of the function invocation}
    \State \textbf{global} $invId$
    \State $invId \gets iid$
  \EndFunction

  \Statex
  \Function{onInvokeFun}{iid, f, base, args, result}
    \State \textbf{global} $L_D$, \StackC{} 
    \If{\Call{iidToLocation}{iid} $\in L_D$} \Comment{Dependency function/method}
      \If{\Call{getActiveCtxType}{} = CMP} \Comment{Invoked in the context of target component}
        \State $ctx_t \gets$ \Call{findTestContext}{\StackC{}}
        \State $t \gets \Call{func}{ctx_t}$ \Comment{Get the func member of the tuple}
        \State $T_C \gets T_C \cup \{t\}$
      \EndIf
    \EndIf
  \EndFunction

  \Statex
  \Function{onFunctionEnter}{iid, func, receiver, args}
    \State \textbf{global} $invId$, \StackC{}
    \State $\Call{activateExecutionContext}{iid, invId, func}$
  \EndFunction

  \Statex
  \Function{onFunctionExit}{iid, returnVal}
    \State \textbf{global} \StackC{}
    \State \Call{pop}{\StackC{}}
  \EndFunction

\end{algorithmic}
\end{algorithm}

The management of the \StackC{} takes place in hook functions
executed before and after the invocation of functions
declared in the project's source code and called during dynamic analysis.
Their implementation is described in pseudocode in Algorithm~\ref{alg:test-filter}.
Specifically, the \code{onFunctionEnter(iid, func, receiver, args)} hook (lines 20--23)
executes before a function invocation 
and creates its correspoding execution context tuple at the top of \StackC{}.
The execution context tuple is removed from \StackC{}
by the \code{onFunctionExit(iid, returnVal)} hook (lines 24--26),
which is invoked after the completion of each function invocation.

More specifically, \code{onFunctionEnter} 
receives as arguments the $iid$ of the function declaration,
a reference $func$ to the function declaration object,
a reference $receiver$ to the receiver object in case of method invocation,
and a list $args$ of the actual arguments of the invocation.
Its implementation has access to the global variable $invId$ 
that stores the iid of the latest function invocation instruction.
The AST location of $invId$ corresponds to the call site 
that activated the function enter hook.
The $invId$ global is defined in the \code{onInvokeFunPre} hook,
executed before any function invocation (lines 6--9).
The function enter hook invokes the utility function 
\code{activateExecutionContext(iid, invId, func)} which is presented,
along with other utility functions, in Algorithm~\ref{alg:ctx-mgmt}.
Its implementation (lines 1--11) determines the execution context type,
through matching the iid of its invocation statement 
against the locations in $L_T$, $L_D$.
Then, it creates the respective execution context tuple and
sets its parent context to the current top of \StackC{} .
The new execution context is pushed at the top of the stack
to serve as parent context of all function invocations 
spawned within its statements (lines 9--11).
Finally, its is removed from the top of \StackC{},
on function completion, by the \code{onFunctionExit} hook.

Given the stack of execution contexts,
test case filtering takes place after each function invocation 
as part of the \code{onInvokeFun} hook (Algorithm~\ref{alg:test-filter}, lines 10--19).
Specifically, each time the invocation instruction's iid corresponds to a component dependency call site (line 12),
the algorithm gets the type of its parent execution context,
which is located at the top of \StackC{},
with a call to the utility function \code{getActiveCtxType}.
In case that the invocation originated from the target component $C$
(context type equals to $CMP$) 
the call stack is, further, analyzed
to identify the respective test case execution context
and, finally, the test method $t$ that reaches the invocation (lines 13--15).
The traversal of \StackC{} is performed by \code{findTestContext},
which represents a simple list scanning algorithm, not included for brevity reasons.
Predicate \code{func($ctx_t$)} represents
the extraction of $func$ item from the $ctx_t$ tuple (line 15).
Matched test methods are added to the set $T_C$
that is returned as algorithm output (line 16).

\begin{algorithm}
\caption{\enskip Execution context management functions}\label{alg:ctx-mgmt}
\scriptsize
\begin{algorithmic}[1]

  \Function{activateExecutionContext}{iid, invId, func}
    \State \textbf{global} $loc_C$, $L_T$, $L_D$, \StackC{}
    \If{$\Call{iidToLocation}{invId} \in L_T$} $type \gets TEST$
    \ElsIf{$\Call{iidToLocation}{invId} \in L_D$} $type \gets DEP$ \Comment{Dependency function/method}
    \ElsIf{$\Call{belongsToAst}{loc_C, iid}$} $type \gets CMP$ \Comment{Target component function/method}
    \Else 
      \State $type \gets OTHER$
    \EndIf
    \State $ctx_p \gets$ \Call{top}{\StackC{}}
    \State $ctx \gets (invId, func, type, ctx_p)$
    \State \Call{push}{\StackC{}, $ctx$}
    \State \textbf{return} $ctx$
  \EndFunction

  \Statex
  \Function{getActiveCtxType}{}
        \LineComment{Returns the type of the currently active context}
        \State \textbf{global} \StackC{}
        \State $ctx \gets$ \Call{top}{\StackC{}}
        \State \textbf{return} $\Call{type}{ctx}$
  \EndFunction

\end{algorithmic}
\end{algorithm}

\subsection{Identification of seed execution paths} 
\label{sect:seed-paths}

This step involves dynamic analysis of the filtered test cases
in order to trace the execution paths
that cover the call sites of dependencies $d_i \in D$.
These execution paths will be referred to as ``seed'' execution paths,
since they seed the test generation process
with essential information for the production of test fixtures and assertions.

A seed execution path $p$
comprises a sequence of statements $(n_0, n_1,\ldots, n_t)$,
starting with the first statement $n_0$ of a test case $t$,
and terminating to the last statement $n_t$ of the test case.
A seed execution path includes at least one statement
that directly invokes a dependency $d_i \in D$, i.e. a call site of $d_i$.
All statements in a seed path must be declared in the implementation 
of the test case $t$ or the target component $C$.

Each statement $n$ in a seed path is represented 
as a tuple $(iid_n, ctx_n, V_n, R_{u,n}, R_{d,n}, CTX_{n})$, where
(a) $iid_n$ is an iid characterizing the AST location of the statement,
(b) $ctx_n$ is the execution context of the statement,
(c) $V_n$ is a set $(name, value)$ pairs,
where $name$ represents a variable defined to a primitive value within statement $n$
and $value$ is the actual value,
(d) $R_{u,n}$ represents a set of object references 
used by the statement,
(e) $R_{d,n}$ represents the set of object references
mutated by the statement, i.e. one or more of their properties 
are defined to a new value, and
(f) $CTX_n$ is a list of execution contexts 
spawned by statement $n$ due to function invocations.
For the purposes of seed path identification,
each execution context instance $ctx_i$ is extended 
with the following attributes:
(i) the list $S_i$ of statements executed as part of the 
context's respective function invocation,
(ii) a reference $rcv_i$ to the receiver object of the method invocation (if any),
(iii) the list of argument values $args_i$ provided to the function invocation,
(iv) the return value of the function invocation $res_i$.

\begin{figure}[hbt!]
    \centering
    \captionsetup[figure]{justification=centering}
    \includegraphics[scale=0.7]{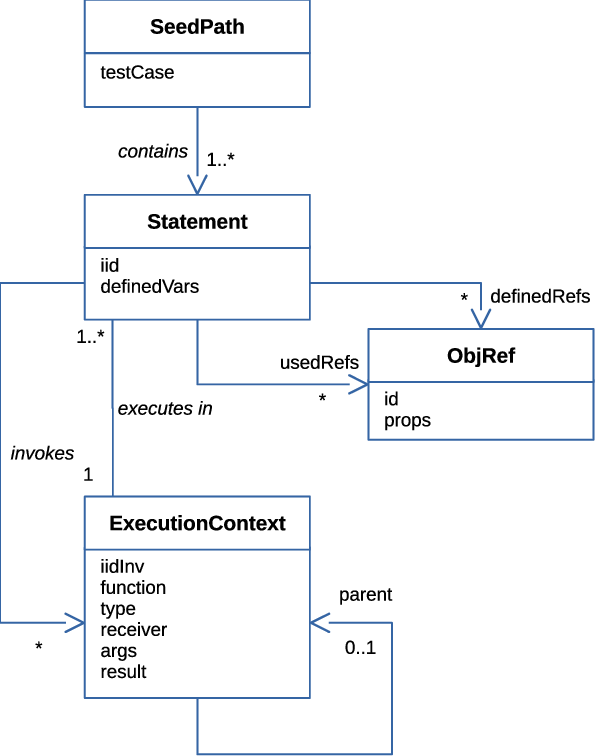}
    \caption{Domain model of dynamic analysis concepts.}
    \label{fig:dm}
\end{figure}

During our analysis, 
we abstract object references with \ObjRef{} objects,
i.e. plain objects that represent the identity 
of an object created during application execution.
Each application object reference is mapped to a unique \ObjRef{} instance.
An \ObjRef{} instance $o_k$ is represented as a tuple $(id_k, props_k)$,
where $id_k$ corresponds to a unique id assigned to it upon creation.
The $id$ is not depended in any way to the $iid$ of any statement
and has its own value range.
The $prop_k$ field is a set of pairs $(prop, value)$,
where $prop$ represents the name of an object property 
assigned to the application object that $o_k$ reference abstracts,
and $value$ is a primitive value assigned to that property
or another \ObjRef{} instance in case that the property 
was defined to a reference value.
For performance reasons, 
the $props_k$ is populated only for objects that are mutated
within the execution context of component dependencies.
The cached property values will be employed 
during construction of the assert part of the generated test cases. Figure~\ref{fig:dm} presents a class diagram 
that represents the concepts involved in seed path identification
and their structure in terms of attributes
and relationships.

The proposed algorithm maintains a global map $allRefs$
of references to \ObjRef{} instances,
for each object reference 
that is valid during dynamic analysis.
The map implementation is platform dependent 
and allows garbage collection (e.g. \texttt{WeakMap} in Node.js).
Function \texttt{getObjRef} in listing~\ref{alg:instruction-utils} facilitates 
the retrieval of the corresponding \ObjRef{} for a given value.
For each object used within an instruction 
the respective \ObjRef{} instance is appended to the set $allRefs$.
Note that the proposed algorithm does not track references
to platform built-in objects (e.g. \code{String}, \code{Date})
which are treated as values.
Instead it focuses on used-defined objects
that are identified by the predicate \code{isObject(val)}.
Due to its simplicity a detailed algorithm for \code{isObject(val)}
is not provided.

\begin{algorithm}
\caption{\enskip Utility functions in seed path identification}
\label{alg:instruction-utils}
\scriptsize
\begin{algorithmic}[1]

  \Statex 
  \Function{getObjRef}{value}
    \LineComment{Returns a new or existing ObjRef instance for the given value}
    \Global{$allRefs$}
    \If {$! \Call{isObject}{value}$} 
      \Returns{$null$}
    \EndIf
    \State $objRef \gets \Call{get}{allRefs,value}$
    \If {$objRef = null$} 
      \State $objRef \gets new ObjRef()$
      \State $\Call{set}{allRefs, value, objRef}$
    \EndIf

    \Returns{$objRef$}
  \EndFunction

\end{algorithmic}
\end{algorithm}

The seed path identification algorithm analyzes 
statements, instructions and function execution contexts 
during dynamic analysis and its steps are executed
as part of appropriate hook invocations.
The algorithm receives as input 
(a) a list of AST locations $L_T$ corresponding to the filtered test cases,
(b) a list of AST locations $L_D$ for all call sites of 
target component dependencies,
(c) the AST location $loc_C$ of the target component $C$.
It returns a list of seed paths $P$ with all required information
for the generation of test fixtures and assertions.

\begin{algorithm}
\caption{\enskip Seed Path Identification}\label{alg:seed-paths}
\scriptsize
\begin{algorithmic}[1]
  \State INPUT: $L_T$, $L_D$, $loc_C$
  \LineComment{Set of seed execution paths}
  \State OUTPUT: $P$ 
  \State $P \gets \emptyset$
  \LineComment{Current seed path}
  \State $P_a \gets ()$

  \LineComment{Other initializations for context management}
  \Statex $\ldots$

  \Statex
  \Function{startStatement}{iid, type}
    \State \textbf{global} $P_a$, \StackC{}, \StackS{}
    \State $ctx_n \gets$ \Call{top}{\StackC{}}
    \State $ type \gets \Call{type}{ctx}$
    \If{$type \in \{DEP, OTHER\}$} $\textbf{return}$ \EndIf

    \State $n \gets (iid_n, ctx_n, \emptyset, \emptyset, \emptyset)$
    \State $P_a \gets P_a \cdot n$
    \State $\Call{addStatement}{ctx_n, n}$
    \State \Call{push}{\StackS{}, $n$}
  \EndFunction

  \Statex
  \Function{endStatement}{iid, type}
    \State \textbf{global} \StackS{}
    \State \Call{pop}{\StackS{}}
  \EndFunction

  \Statex
  \Function{onFunctionEnter}{iid, func, rcv, args}
    \State \textbf{global} $invId$, \StackC{}, \StackS{}
    \State $ctx \gets \Call{activateExecutionContext}{iid, invId, func, receiver, args}$
    \State $ type \gets \Call{type}{ctx}$
    \If{$type = TEST$} \Comment{Entering a new test function}
        \State $P \gets P \cup \{P_a\}$
        \State $P_a \gets ()$
    \EndIf
    \State $n \gets$ \Call{top}{\StackS{}}
    \State $\Call{addExecutionCtx}{n, ctx}$
  \EndFunction
  
  \Statex
  \Function{onFunctionExit}{iid, returnVal}
    \State \textbf{global} \StackC{}
    \State $ctx \gets$ \Call{pop}{\StackC{}}
    \State $\Call{addResult}{ctx, returnVal}$
  \EndFunction

\end{algorithmic}
\end{algorithm}

The algorithm analyzes runtime constructs of the instrumented code
and operates at three levels of analysis 
(i) statement, (ii) function and (iii) instruction. 
A pseudocode description of the algorithm, 
specifying its steps as part of dynamic analysis hooks,
is presented in Algorithm~\ref{alg:seed-paths}.
The three levels of dynamic analysis are outlined below.\\

\textbf{Statement level}: the execution of a new statement $n$
is signalled by the invocation of the \code{startStatement(iid, type)} hook.
The hook is invoked prior to statement execution, 
while a matching \code{endStatement(iid, type)} is invoked after its execution.
On statement entry the algorithm peeks the current execution context $ctx_n$
from the \StackC{} and checks its type (lines 8--10).
In case of \code{DEP} or \code{OTHER} context type
statement processing is aborted, since test fixture generation 
does not take into account statements declared
outside the test cases in $L_T$ or the target component $C$.

The start statement hook creates a new statement tuple 
$n = (iid_n, ctx_n, \emptyset{}, \emptyset{}, \emptyset{})$
and appends it to the current seed path,
accessible through the global variable $P_a$ (lines 11--12).
Statement $n$ is, also, appended to the statement list $S_n$ 
of the current execution context $ctx_n$.
The sets $R_{u,n}$, $R_{d,n}$ and $CTX_n$ are populated
from other hooks that will be specified hereafter.
In order to track the execution contexts activated by $n$,
the algorithm maintains a statement stack \StackS{},
where statements are pushed in the \code{startStatement} hook (line 14)
and popped in the \code{endStatement} (line 18).
Note that statement execution may involve invocations 
of one or more functions declared in the current test case
or the target component $C$.
Each function executes one or more statements,
which in their turn may invoke other functions etc.
The statement stack allows upon function entry
to trace the invoking statement which is located at the top of the stack.\\

\textbf{Function execution level}: the execution of a function
that belongs to the analyzed project is framed by 
\code{functionEnter(iid, f, receiver, args)} 
and \code{functionExit(iid, result)} hooks
that implement context management functionality,
as specified in the previous section.
However, in this step, each execution context instance
is supplemented with the $receiver$ and $args$ information (line 22),
as well as the returned $result$ 
which is available on function exit (lines 33-34).
The utility function \code{addResult} updates the current context tuple $ctx$
with the return value of the recently completed function.

Besides context management,
the \code{functionEnter} hook handles seed path creation
and updates the content of the set $P$ of all identified seed paths.
Specifically, in case that the currently created context is $TEST$,
i.e. a new test function is activated,
the hook updates the set $P$ with the current seed path $P_a$
and resets $P_a$ to an empty path (lines 23--27).
Moreover, \code{functionEnter} populates 
the list of contexts $CTX_n$ spawned by the current statement $n$,
i.e. the statement that invoked function $func$.
The current statement is available at the top of \StackS{}
and $ctx$ is added to its context list $CTX_n$ 
with the help of utility function \code{addExecutionCtx()}~(lines 28--29).\\

\textbf{Instruction level}: each statement involves 
the execution of one or more instructions,
e.g. read/write of variables or object properties,
function invocations, creation of object literals.
Instruction execution is signalled by appropriate hooks
that are invoked before and after its evaluation.
The proposed algorithm harnesses ``after'' hooks
in order to track used and mutated object references
for the current statement $n$,
which is peeked from \StackS{}.
Moreover, it tracks updates of object properties with primitive values 
as part of component dependency invocation statements.
Updated properties will set the basis 
for the construction of the assert part of the generated test case.
Involved hooks and their role is described below:

\begin{itemize}
  \item \code{read(iid, name, val)} is invoked after instruction $iid$,
  reading value $val$ from variable $name$.
  In case that \code{isObject(val)} is true,
  an \ObjRef{} instance is retrieved 
  and added to the set of used references $R_{u,n}$
  for current statement $n$.
  
  \item \code{write(iid, name, val)} is invoked after instruction $iid$,
  writing value $val$ to variable $name$.
  The \ObjRef{} instance that corresponds to $val$ is added to the set $R_{d,n}$, 
  once \code{isObject(val)} evaluates to true.
  In case that $val$ corresponds to a primitive type,
  the pair $(name, val)$ is added to the set of defined variables $V_n$
  of the current statement.

  \item \code{getField(iid, base, offset, val)} executes after instruction $iid$,
  reading a value $val$ from property named $offset$ of object reference $base$.
  In case that $base$ and $val$ are user-defined objects,
  their respective \ObjRef{} instances is added to the set $R_{u,n}$.

  \item \code{putField(iid, base, offset, val)} is invoked after instruction $iid$,
  assigning value $val$ to the $offset$ property of object $base$.
  The value in $val$ is treated in the same way as in \code{getField}.
  However, since the referenced object in $base$ is mutated,
  its respective \ObjRef{} is added to the set $R_{d,n}$ of current statement $n$.
  Especially, in cases that the execution context of the current statement
  is spawned by a component dependency 
  (current or parent execution context of type $DEP$),
  the pair $(offset, val)$ is added to the $props$ set of $base$ \ObjRef{}.
  In case that $val$ is an object reference, it's respective \ObjRef{} is used instead.
  Mutated properties within invocations of component dependencies
  will set the base for generation of test case assertions.

  \item \code{invokeFun(iid, f, base, args, result)} executes
  after the invocation of function $f$ in instruction $iid$.
  The function invocation arguments are available in the list $args$,
  while $result$ contains the returned value.
  In case that $base$ is present, the invocation corresponds
  to a method call on the object referenced by $base$.
  All user-defined objects referenced by $base$, $val$ and $args$
  are mapped to \ObjRef{} instances 
  and added to the set of used references $R_{u,n}$
  for current statement $n$.
  
  \item \code{literal(iid, val)} corresponds to the creation 
  of a literal $val$ in instruction $iid$. 
  In case that $val$ is a user-defined object,
  i.e. \code{isObject(val)} is true,
  the respective \ObjRef{} instance is added to the set of defined references $R_{d,n}$
  for current statement $n$.

\end{itemize}

\subsection{Tracing object flow dependencies in seed execution paths}
\label{sect:data-flow}

In this step, we trace the object flow dependencies
among the statements of seed execution paths.
Let $p$ be a seed execution path 
and $n_a$, $n_b$ are statements belonging to $p$,
with $n_a$ preceding $n_b$ in execution sequence.
We define an \emph{object flow dependency} as a pair $(n_b, n_a)$,
where statement $n_b$ uses at least one object 
whose state is mutated in statement $n_a$.
In other words, there exists at least one \ObjRef{} instance $o_i$
with $o_i \in R_{u,b} \wedge o_i \in R_{d,a}$,
ie. the object reference is simultaneously contained
in the set of used objects of $n_b$ and the set of defined objects of $n_a$.
Object flow dependencies within a seed execution path 
enable the extraction of backward program slices,
having as seed statements the call sites of dependencies in $D$.
The statements of a backward slice are appropriately modified,
on the basis of program state collected during dynamic analysis,
in order to generate the code required for test fixture setup.

The algorithm for tracing object flow dependencies 
within a given seed execution path is presented in Algorithm~\ref{alg:obj-flow}.
The algorithm receives as input a seed path $p_t = (n_0, \ldots, n_t)$, 
as well as the set $L_D$ of call sites for component dependencies in $D$.
It returns as output a set $F_t$ of object flow dependencies
among the statements of $p_t$.
The algorithm is applied iteratively to all seed paths $P$,
identified during dynamic analysis,
and complements all required program state 
to drive the next stage of test generation.

\begin{algorithm}
\caption{\enskip Object Flow Dependencies Tracing}\label{alg:obj-flow}
\scriptsize
\begin{algorithmic}[1]
  \State INPUT: $p_t: (n_0, \ldots, n_t), L_D$
  \State OUTPUT: $F_t$ 
  
  \State $F \gets \emptyset$
  \State $k \gets 0$
  \LineComment{Locate the last call site of component dependencies}
  \For{$i \gets t$ \textbf{to} 0}
    \State $iid_i \gets \Call{iid}{n_i}$
    \If{$\Call{iidToLocation}{iid_i} \in L_D$}
      \State $k \gets i$
      \State \textbf{break}
    \EndIf
  \EndFor
  \Statex
  \LineComment{Backward list traversal to identify object flow dependencies}
  \For{$i \gets k$ \textbf{to} 1}
    \State $R_{u,i} \gets \Call{usedRefs}{n_i}$
    \For{$j \gets i-1$ \textbf{to} 0}
      \State $R_{d,j} \gets \Call{definedRefs}{n_j}$
      \If{$R_{u,i} \cap R_{d,j} \neq \emptyset$}
        \State $F_t \gets F_t \cup \{(n_i, n_j)\}$
      \EndIf
    \EndFor
  \EndFor

\State \textbf{return} $F_t$
\end{algorithmic}
\end{algorithm}

Tracing of object flow dependencies involves backwards traversal
of the list $p_t$ starting from statement $n_t$.
The analysis begins with the first encountered statement $n_k$
that contains a call site to a component dependency in $D$.
Statements following $n_k$ do not contribute to test fixtures
and, thus, are not processed.
Lines 5--10 in Algorithm~\ref{alg:obj-flow} identify the last index $k$, 
such that the $iid$ of the statement $n_k$ matches
an AST location in the set $L_D$ of target component dependencies' call sites.
Then, starting from $n_k$, each statement is evaluated against its predecessors
in a double nested loop to check for object flow dependencies (lines 12--20).
Specifically, function \code{usedRefs()} returns the set of used \ObjRef{} instances,
$R_{u,i}$, for each statement $n_i$ (line 13),
which is intersected with the set of defined \ObjRef{} instances $R_{d,j}$
of predecessor statements $n_j, j \in \{0,\ldots,i-1\}$ (lines 14--19).
The set $R_{d,j}$ is retrieved from tuple $n_j$ 
through invocation of function \code{definedRefs()}.
In case that the intersection $R_{u,i} \cap R_{d,j}$ is not empty,
a new object flow dependency from statement $n_i$ to statement $n_j$
is appended to the set $F_t$ (lines 16--18).

\begin{figure}[hbt!]
    \centering
    \captionsetup[figure]{justification=centering}
    \includegraphics[scale=0.7]{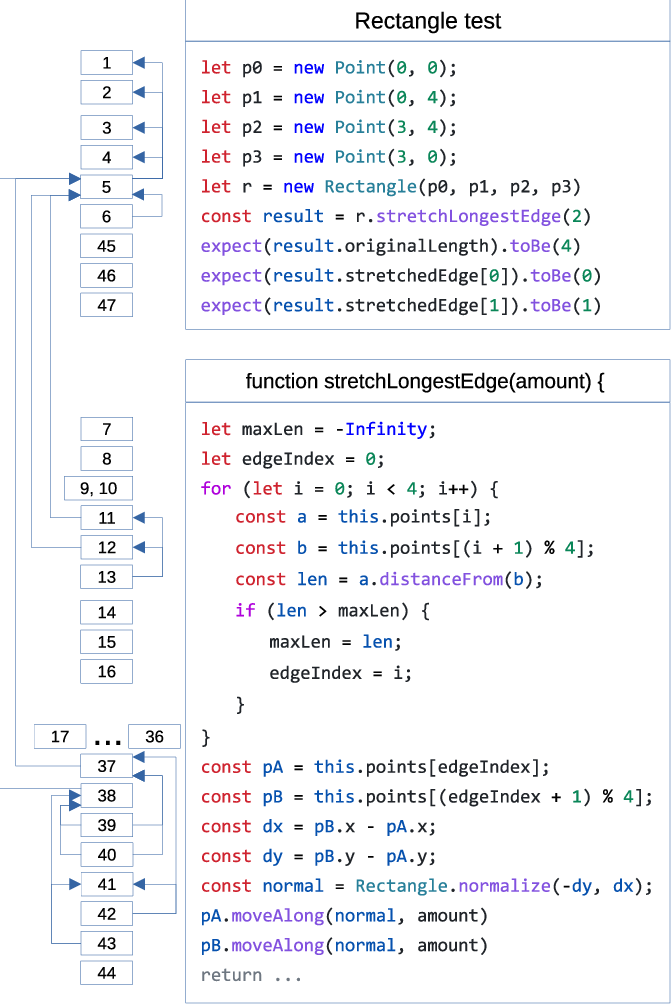}
    \caption{Object flow dependencies for the seed path of the case study integration test.}
    \label{fig:obj-flow}
\end{figure}

Figure~\ref{fig:obj-flow} illustrates 
the object flow dependencies traced within the seed path
corresponding to the execution of the integration test for \code{stretchLongestEdge()}.
The seed execution path comprises 47 statements 
belonging to the implementation of the test case and the \code{stretchLongestEdge()}.
Its length outnumbers the statements of the test case and the target component,
since it involves 4 iterations of a \code{for} loop
that starts at statement 9 and terminates at statement 36.
Notice that the nested conditional in the \code{for} loop is executed only once.
Tracing of object flow dependencies starts at statement 43
which involves the last invocation of a component dependency (\code{moveAlong() method}).
More specifically, statement 43 uses objects referenced by \code{normal} and \code{pB}
that are defined in statements 41 and 38, respectively,
resulting to object flow dependencies $(43, 41)$ and $(43, 38)$.
Tracing of object flow dependencies for statements 5, 6, 13, 39, 40, 42 
follows the same logic.
Attention should be drawn to statements of the target component
that reference \code{this} and contribute to interprocedural dependencies.
For instance, the \code{this} reference in statement 38 points 
to a \code{Rectangle} object, abstracted by the same \ObjRef{} instance 
with that defined in statement 5,
Thus, an object flow dependency $(38, 5)$ is added among the respective statements.
The same applies to statements 11, 12, 37 that, also, reference \code{this}.

\subsection{Test case generation}
\label{sect:test-generation}

In this step, each seed execution path,
enhanced with object flow dependencies among its nodes,
is processed for the generation of test cases.
The proposed method produces one test case 
for each call site of a dependency in $D$,
included in the seed execution path.
Test generation processes (a) the program state collected during dynamic analysis and
(b) AST information of the integration test and target component implementations
in order to construct the statements of the essential parts of a test case:

\begin{itemize}
  \item \emph{Arrange part}: statements that setup the test fixture
  \item \emph{Act part}: invocation of the method under test
  \item \emph{Assert part}: assertion statements that verify the test outcome
\end{itemize}

Assume that $AST_t$, $AST_C$
be the AST representations of an integration test $t$
and the target component $C$.
Let $p_t = (n_0, n_1,\ldots, n_t)$ be a seed execution path
corresponding to the execution of the test case $t$.
The test case generation algorithm receives 
$AST_t$, $AST_C$, $p_t$ as input,
and returns a set $L_G$ comprising the AST representations
of generated test cases.
The steps of the test generation algorithm are outlined below:

\begin{enumerate}[label=S\arabic*.]
  \item Locate the next statement position in $p_t$, starting from 0, 
  that corresponds to a component dependency call site. 

  \begin{enumerate}[label=(\alph*)]
    \item If no call site is identified, return $L_G$.
    \item Let $k$ be the position a dependency call site and $n_k$ the respective statement.
  \end{enumerate}
  
  \item Generate the \emph{Arrange} and \emph{Act} parts of the test case.
  \begin{enumerate}[label=(\alph*)]
    \item Invoke algorithm \textsc{generateArrangeAct} with input $AST_t$, $AST_C$, $p_t$, $F_t$, $n_k$.
    \item Store the algorithm output to list $l_a$ of statements.
  \end{enumerate}

  \item Generate the \emph{Assert} part of the test case.
  \begin{enumerate}[label=(\alph*)]
    \item Invoke algorithm \textsc{generateAssert} with input $AST_t$, $AST_C$, $p_t$, $n_k$.
    \item Store the algorithm output to list $l_b$, comprising generated statements for the Assert part.
    \item Concatenate lists $l_a$ and $l_b$ to the full statement list $l_k$ for the generated test case.
  \end{enumerate}

  \item Generate the complete $AST_k$ of the test case on the basis of $l_k$.
  \begin{enumerate}[label=(\alph*)]
    \item Set $L_G \gets L_G \cup \{AST_k\}$
  \end{enumerate}

  \item Continue execution from \emph{Step 1} and starting position $k+1$.
\end{enumerate}

The algorithm is iteratively applied to all integration tests in $L_T$ 
and their respective seed paths in $P$ to produce the final augmented test suite.
In the rest of this section, we provide more detailed descriptions
for algorithms \textsc{generateArrangeAct} and \textsc{generateAssert}. 

\subsubsection{Generation of the Arrange and Act statements}

The algorithm that generates the Assert and Act parts for a unit test
that exercises a specific call site of a component dependency,
receives as input: 
(a) the AST representations $AST_t$ and $AST_C$ 
of the integration test and the target component,
(b) the seed path $p_t = (n_0, n_1,\ldots, n_t)$ 
corresponding to the execution of the integration test,
(c) the set of object flow dependencies $F_t$ for path $p_t$,
(d) a representation $n_k = (iid_k, ctx_k, V_k, R_{u,k}, R_{d,k}, CTX_{k}), k \leq t$
of the statement that corresponds to the call site.
For brevity reasons, 
we will refer to the target component invocation statement as $n_c \in p_t, c < k$. 
Statement $n_c$ can be easily identified within $p_t$ 
through locating the statement that spawns the context of $n_k$,
i.e. $ctx_k \in CTX_{c}$.

The algorithm returns a list $l_a$ with the AST representations 
of the generated statements.
Starting from statement $n_k$ in position $k$ of the seed path
the algorithm follows object flow dependencies 
and selects statements from $p_t$ to set the basis of a backward program slice 
with $n_k$ as the seed statement.
The process is repeated for $n_c$ to complement the slice 
with essential state for the invocation of the target component.
Then, appropriate ``projection'' of the AST representation of $n_k$ to a simplified AST,
on the basis of program state collected during dynamic analysis,
results to the final form of the Act statement.
Similar processing of the rest of the backward slice
produces the statements of the Arrange part of the unit test.

\paragraph{AST Projection}

Let $ast_i$ be the AST representation of the statement $n_i$,
which is retrieved from $AST_t$ or $AST_C$
based on location information returned by \code{iidToLocation()}.
The AST projection of statement $n_k$, or any other statement $n_i$,
is returned by function \code{projectAST($n_k$, $ast_k$, $thisId$)} 
that receives as arguments the AST representation $ast_k$ of the statement,
its program state $n_k = (iid_k, ctx_k, V_k, R_{u,k}, R_{d,k}, CTX_{k})$
and an identifier to replace all occurrences of the \code{this} reference.
The function returns a simplified AST representation for $n_k$, $ast_k'$,
where selected expressions of function calls
are replaced by their runtime values.
Moreover all references to \code{this} are replaced with identifier $thisId$.
An outline of the \code{projectAST($n_k$, $ast_k$, $thisId$)} processing is provided below:

\begin{enumerate}[label=S\arabic*.]
  \item Create a copy $ast_k'$ of $ast_k$ to apply modifications
  \item If $thisId$ is not empty, 
    replace all \code{ThisExpression} occurrences in $ast_k'$
    with an \code{Identifier} object named $thisId$
  \item Traverse $ast_k'$ breadth first, 
    in search of AST nodes with \code{CallExpression} type

  \item Let $c_i$ be the next visited \code{CallExpression} node during $ast_k'$ traversal
  \item Locate the execution context $ctx_i \in CTX_k$,
    corresponding to the AST location of call expression $c_i$
  \begin{enumerate}[label=(\alph*)]
    \item Let $args_i$ be the list of actual argument values,
    stored in $ctx_i$
  \end{enumerate}

  \item For each expression $e_j$ contained in the \code{arguments} attribute of AST node $c_i$
  do the following:
  
  \begin{enumerate}[label=(\alph*)]
    \item Locate the respective value $v_j$ in actual arguments of $ctx_i$
    \item If $v_j$ is an \ObjRef{} instance leave the expression $e_j$ intact
    \item Else, analyze the $e_j$ subtree for \code{CallExpression} nodes
    or nodes of type \code{UpdateExpression} / \code{AssignmentExpression} 
    that apply on \code{MemberExpression} nodes
    \begin{enumerate}[label=(\roman*)]
      \item if any node is identified, leave $e_j$ intact
      \item Else, replace $e_j$ with a \code{Literal} node of value $v_j$
    \end{enumerate}
  \end{enumerate}
  \item Continue execution from S4 with $ast_k'$ traversal
  \begin{enumerate}[label=(\alph*)]
    \item If no \code{CallExpression} node is located 
    \emph{return} $ast_k'$ and terminate
  \end{enumerate}
\end{enumerate}

\begin{figure}[hbt!]
    \centering
    \captionsetup[figure]{justification=centering}
    \includegraphics[scale=0.7]{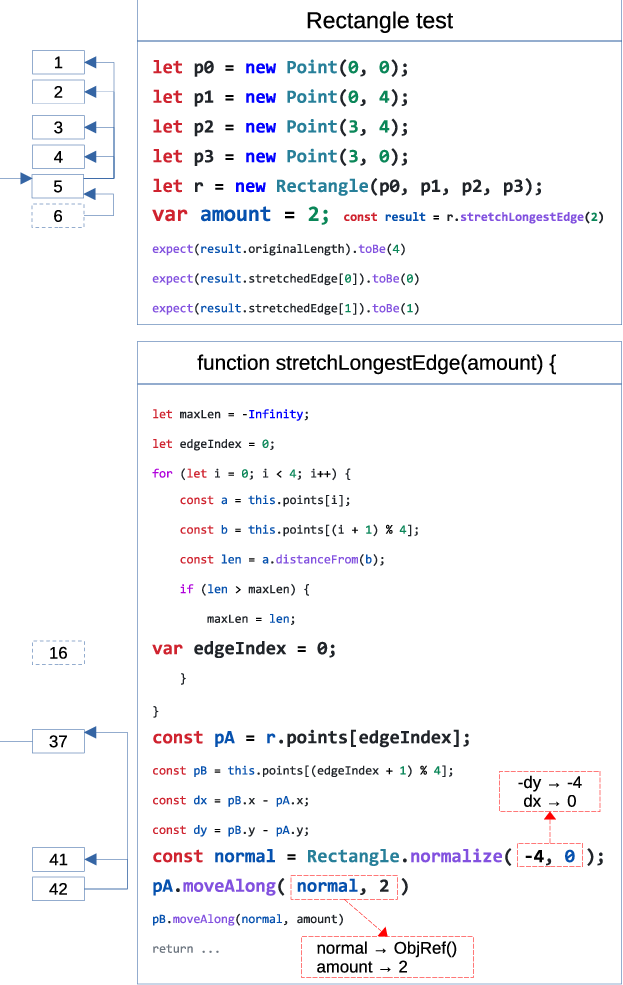}
    \caption{Generation of the Arrange-Act parts of the unit test for call site in statement 42.}
    \label{fig:arrange-act}
\end{figure}

The intuition behind the AST projection is that expressions
provided as arguments to \code{CallExpression} nodes
within the AST of a statement $n_k$ can be replaced
with actual argument values collected during dynamic analysis,
resulting to a simplified statement during code generation.
Thus, the simplified statement has less dependencies to program variables
and the size of the backward slice for test fixture setup is minimized.
However, a requirement is to preserve expressions that mutate object state,
e.g. expressions referred to step S6(c).
These expressions are required since, for performance reasons,
the proposed method does not store object state during dynamic analysis 
but, instead, relies on test fixture code to reinstate it.
For instance, in Figure~\ref{fig:arrange-act},
the AST projection for statement 42 of the seed path
results to statement \code{pA.moveAlong(normal, 2)}.
The runtime value of the first argument is an object reference,
while the second argument was replaced by its actual value.
On the other hand, the projection of statements 37, 38 
does not affect their original AST representation,
since they do not include any \code{CallExpression} nodes.

Notice that AST representations returned by \code{projectAST()} include
identifiers corresponding to variable names.
These identifiers should be defined to their runtime value, for the given execution path,
through appropriate assignment statements.
The extraction of pending identifiers from an $ast_k'$ is performed 
through function \code{extractIdentifiers(ast)} 
that returns a list of identifiers to be defined.
For instance, the returned identifiers for statement 37 are \code{[this, edgeIndex]},
while for statement 42 are \code{[pA, normal]} (see Figure~\ref{fig:arrange-act}).

\paragraph{\textsc{GenerateArrangeAct} Algorithm}

An outline of the \textsc{GenerateArrangeAct} algorithm is provided below:

\begin{enumerate}[label=S\arabic*.]
  \item Locate statement $n_c \in p_t$ corresponding to target component invocation 
  \item Resolve \code{this} reference to an identifier of the parent scope
  through invocation of \code{resolveThis($n_k$)}
  \begin{enumerate}[label=(\alph*)]
    \item Store the identifier to variable \code{thisId} 
  \end{enumerate}

  \item Traverse object flow dependencies in $F_k$ starting from statements $n_k$, $n_c$
  \begin{enumerate}[label=(\alph*)]
    \item Collect all reachable statements, including $n_k$, $n_c$ to list $N_k$
    \item Sort list in descending order of statement appearance in $p_t$
  \end{enumerate}

  \item Initialize a list of null values $l_a$ with size $k+1$ 
  to store the ASTs of statements in the backward slice from $n_k$
  \item For each $n_i \in N_k$ do 
  \begin{enumerate}[label=(\alph*)]
    \item Set $ast_i' \gets \code{projectAST}(n_i, ast_i, thisId)$
    \item Set $idx$ to the position of $n_i$ in seed path $p_t$
    \item Store $ast_i'$ to the position $idx$ in list $l_a$
  \end{enumerate}

  \item Replace $ast_c \in l_a$ with a \code{VariableDeclaration} statement
   produced by $\code{getParamsDeclaration}(n_c, l_a)$

  \item Traverse list $l_a$ backwards starting from last item \label{step:exportPending}
  \begin{enumerate}[label=(\alph*)]
    \item Store next \emph{not null} item to variable $ast_i$,\label{step:nextAst}
    where $i$ is statement position in $l_a$
    \item If end of list is reached, \emph{return} $l_a$ as algorithm result and terminate.
  \end{enumerate}

  \item Extract pending identifiers for $ast_i$, $I_i \gets \code{extractIdentifiers}(ast_i)$
  \item For each $id_j \in I_i$ do
  \begin{enumerate}[label=(\alph*)]
    \item Search $p_t$ backwards, starting from position ${i-1}$
    for the first item $n_m$ having $id_j$ in the set of defined variables $V_m$
    \begin{enumerate}[label=(\roman*)]
      \item If no statement found, \emph{continue} with next id in $I_i$\\
       ($id_j$ references an object or parameter and its statement is already in $l_a$)
      \item If the same index item $ast_m$ in $l_a$ is not null, 
      \emph{continue} with next id in $I_i$ \\
      (statement, already, generated due to processing of a predecessor item in $l_a$)
      \item Let $(id_j, v_j) \in V_m$ and $v_j$ the assigned value to $id_j$ in statement $n_m$
      \item Create a \code{VariableDeclaration} AST of kind \emph{var}
      for initialization of $id_j$ identifier to literal value $v_j$
      \item Assign generated AST to position $m$ in list $l_a$
    \end{enumerate}
  \end{enumerate}

  \item Resume from Step~\ref{step:exportPending}(a) with next statement in $l_a$

\end{enumerate}

The operation of the \textsc{GenerateArrangeAct} algorithm can be divided into two parts.
The first part, comprising steps S1--S6, 
resolves an identifier for the \code{this} expression (step S1), 
if referenced within the target component,
and creates the core of the backward program slice 
that constitutes the arrange and act parts of the generated unit test.
The backward slice is created to include all reachable statements
starting from $n_k$, $n_c$ and following object flow dependencies within seed path $p_t$ (step S3).
Figure~\ref{fig:arrange-act} depicts the backward slice seeded by statement 42
for the case study example.
Its statements are represented as nodes with continuous contour,
connected with directed edges corresponding to object flow dependencies.
The backward slice statements are stored in a list $l_a$,
initialized with \code{null} values (step S4).
Specifically, each statement AST of the backward slice is projected to a simplified version
and inserted in $l_a$ in the same position as its counterpart in the seed path (step S5).
The projected AST for $n_c$, $ast_c$, is replaced by a \code{VariableDeclaration} statement
generated by \code{getParamsDeclaration()} function.
The function removes the invocation of the target component, 
while keeping expressions provided as actual parameters 
as initial values of variables named after the formal parameters of the target component.
Thus, references to the target component parameters in the rest of the slice
are resolved to declared variables and do not result to runtime errors.

The second part of the algorithm makes a backwards traversal of list $l_a$,
starting from position $k$,
and complements the slice with appropriate variable declarations 
for unresolved identifiers contained within each $ast_i \in l_a$ (steps S7--S10).
Specifically, given a projected AST $ast_i$,
the algorithm extracts all identifiers corresponding to variable names,
through the invocation of \code{extractIdentifiers($ast_i$)} (step S8).
Any identifier $id_j \in I_i$ corresponding to a variable with a primitive value
would have a matching statement $n_m, m < i$ 
that includes in its dynamic analysis state a pair $(id_j, v_j) \in V_m$.
Value $v_j$ corresponds to its assigned value right before the execution
of statement $n_i$.
Thus, if $ast_m$ for statement $n_m$ is not already in the slice,
an appropriate \code{VariableDeclaration} statement is inserted in position $m$
that assigns $v_j$ to identifier $id_j$ (step S9).
In any other case, the identifier corresponds 
to either a formal parameter or an object reference and, 
thus, it will be initialized by statements already present in $l_a$.
For instance, statement 37 in Figure~\ref{fig:arrange-act}
references identifiers \code{pA}, \code{r} and \code{edgeIndex}.
Among these identifiers, 
only \code{edgeIndex} receives a primitive value 
and is defined to zero in statement 16, 
right before execution of statement 37.
Thus, an appropriate variable declaration is introduced in position 16 of the $l_a$ list
to complement the slice.
As concerning identifiers \code{pA}, \code{r}, 
their processing in step S9 does not yield any additional statement.
However, their values are defined by statements
that are part of the slice in $l_a$.

\subsubsection{Generation of the Assert part}

The generation of assertions that verify the outcome of the Act statement
is based on the \textsc{generateAssert} algorithm.
As mentioned in Section~\ref{sect:call-site-resolution}
we assume that call sites of component dependencies have the basic form
of a function/method invocation: \\

\code{[(const | var | let | "") result = ] [base.]dependency([arguments])} \\

The \textsc{generateAssert} algorithm receives as input
the statement $n_k = (iid_k, ctx_k, V_k, R_{u,k}, R_{d,k}, CTX_{k})$
that corresponds to the call site,
as well as its AST representation $ast_k$.
The algorithm generates the AST representations of assertion statements
and appends them to a list $l_b$, returned as output.
Assertion statements compare identifiers against their expected value 
and their syntax depends on the preferred testing framework of the user.
For instance, assertions in the popular Jest\footnote{https://jestjs.io} framework
have the form \code{expect(identifier).toBe(value)}.
The AST of an assertion statement is generated 
by \code{genAssertEquals(identifier, value)} function,
whose specification is simple and omitted for brevity reasons.

Let $ctx_i \in CTX_{k}$ be the execution context 
for the exercised component dependency call site,
where $ctx_i = (iid_m, f, type, ctx_k, rcv_i, args_i, res_i)$.
The collected program state for the component dependency invocation includes,
among others, an optional reference to the receiver object $rcv_i$
and the actual returned result $res_i$ which will be assigned to \code{<result>}.
Recall that $res_i$ may correspond to a primitive value 
or an object reference of type \ObjRef{}.
The latter includes a $props$ set of pairs with assigned values 
to properties of the actual runtime object, 
i.e. $props = \{(offset_i, val_i)\}$ where $offset_i$ is the name of a property
and $val_i$ is its respective value.
Notice that $val_i$ might be a primitive value 
or an \ObjRef{} having its own $props$ member.
Context $ctx_i$ can be easily located within the spawned contexts of $n_k$
by comparing the name of the invoked dependency against the referenced function $f$.

The proposed algorithm generates a single assertion statement
in case that $res_i$ corresponds to a primitive value.
If $res_i$ is an object reference 
with one or more defined properties in $props_i$ 
then the algorithm generates one assertion 
for each property assigned to a primitive value.
Properties defined to object references are, 
recursively, handled in the same manner.
Moreover, an analogous set of assertion Statements
is generated for the $rcv_i$ object reference,
should the callee of the dependency invocation is a \code{MemberExpression},
e.g. \code{base.dependency()}.
In order to provide a concise description of the \textsc{generateAssert} algorithm,
we introduce the \code{genObjRefAssertions($prefix$, $objRef$)} utility function.
The function returns a list of assertion statements
corresponding to the defined properties of $objRef$ parameter.
The $prefix$ parameter is introduced as the receiver identifier
to all asserted properties.
The operation of \code{genObjRefAssertions($prefix$, $objRef$)} is provided below:

\begin{enumerate}[label=S\arabic*.]
  \item Initialize an empty list $l$ to store generated assertion statements
  
  \item Let $props$ be the defined properties of the $objRef$ instance
  \item For each $(offset_i, val_i) \in props$
  
  \begin{enumerate}[label=(\alph*)]
    \item If $val_i$ is a primitive value then
    \begin{enumerate}[label=(\roman*)]
      \item Append to $l$ the result of \code{genAssertEquals($prefix.offset_i$, $val_i$)}
    \end{enumerate}
    \item If $val_i$ is an \ObjRef{} instance $obj_i$ then
    \begin{enumerate}[label=(\roman*)]
      \item Append to $l$ the result of 
          \code{genObjRefAssertions($prefix.offset_i$, $obj_i$)}
    \end{enumerate}
  \end{enumerate}
  \item Return $l$ as result
\end{enumerate}

On the basis of all aforementioned definitions, 
an outline of the \textsc{generateAssert} algorithm processing is provided below:

\begin{enumerate}[label=S\arabic*.]
  \item Initialize an empty list $l_b$ to store generated assertion statements
  \item Locate execution context $ctx_i$ for the dependency call site within $CTX_k$
  \item Analyze $ast_k$ to identify the \code{result} and \code{base} AST fragments
  \item If \code{result} is present then
  
  \begin{enumerate}[label=(\alph*)]
    \item If $res_i$ is a primitive value then
    \begin{enumerate}[label=(\roman*)]
      \item Append to $l_b$ the result of $\code{genAssertEquals}(result, res_i)$
    \end{enumerate}
    \item If $res_i$ is an \ObjRef{} instance then
    \begin{enumerate}[label=(\roman*)]
      \item Append to $l_b$ the result of $\code{genObjRefAssertions}(result, res_i)$
    \end{enumerate}
  \end{enumerate}

  \item If \code{base} is present then
  
  \begin{enumerate}[label=(\alph*)]
    \item If $rcv_i$ is a primitive value then
    \begin{enumerate}[label=(\roman*)]
      \item Append to $l_b$ the result of $\code{genAssertEquals}(base, rcv_i)$
    \end{enumerate}
    \item If $rcv_i$ is an \ObjRef{} instance then
    \begin{enumerate}[label=(\roman*)]
      \item Append to $l_b$ the result of $\code{genObjRefAssertions}(base, rcv_i)$
    \end{enumerate}
  \end{enumerate}

  \item Return the list $l_b$ as output
\end{enumerate}

In Figure~\ref{fig:arrange-act}, 
the component dependency invocation in statement 42 does not return any result
and, thus, generated assertions 
verify updates to the state of the receiver object \code{pA}.
Since the \code{moveAlong()} invocation updates both properties of the \code{Point} object,
the respective \ObjRef{} instance representing the state of \code{pA}
contains the set $props = \{(x, -2), (y, 0)\}$.
On the basis of these property definitions 
the algorithm generates two assertions on properties \code{pA.x} and \code{pA.y}
(see Figure~\ref{fig:rectangle-tests}).


\section{Empirical Evaluation} \label{evaluation}

In this section we present an empirical study
of our method for automated unit test generation from existing integration tests.
The \emph{goal} of this study,
as per the goal template specified in~\cite{wohlin2012experimentation}, 
is to analyze the proposed method
for the \emph{purpose} of evaluation
\emph{with respect to} effectiveness and practicality.
The study is performed from the \emph{perspective} of
(a) a researcher that investigates
the degree of test augmentation that can be attained 
by test generation methods,
and (b) a developer that needs to understand 
the potential of the proposed method
in generating tests that are maintainable 
and effective in locating bugs.
The \emph{context} of the study is
a set of open source Node.js applications.

\subsection{Research Questions}

The empirical study aims to answer the following research questions:

\emph{RQ1. What is the degree of test augmentation attained by the proposed method?}
The research question aims to investigate 
the extent at which an existing test suite can be augmented with unit tests.
The results of this research question will provide empirical evidence
on the test generation potential of the proposed method.

\emph{RQ2. Does the proposed method enhance the fault detection capability of a test suite?}
The purpose of this research question is to evaluate 
the effect of generated tests to the fault detection capability of test suites.
The results of this study will highlight 
the potential of the proposed method in generating tests
that are useful not only for regression testing
but, also, for revealing additional faults 
not detected by the original test suite. 

\emph{RQ3. Does the proposed method generate maintainable unit tests?}
The research question focuses on the maintainability of the generated unit tests
Its results will pinpoint the practicality of the proposed method,
in terms of generating tests that are conscise and readable 
so as to be integrated to the codebase without major modifications.

\subsection{Context Selection}

The context of the empirical study comprises twelve Node.js applications. 
The selection of these projects is based on the following criteria: 
(a) they are open source, for study reproducibility reasons, 
(b) they contain components with local implementations of their dependencies, 
(c) they have a test suite which can be successfully executed, 
(d) to the extent possible, they are used as benchmarks in earlier works,
e.g.~\cite{paltoglou2021automated}. 

\begingroup
\setlength{\tabcolsep}{6pt} 
\renewcommand{\arraystretch}{1.2} 
\begin{table}[tb]
\centering
\scriptsize
\begin{tabular}{l r r l}
\toprule
\multicolumn{1}{c}{\textbf{Project}} &
\multicolumn{1}{c}{\textbf{SLOC}} &
\multicolumn{1}{c}{\textbf{Tag}} &
\multicolumn{1}{c}{\textbf{Github Repository}}\\ \hline
node-jsonfile       & 107   & v6.1.0  & {jprichardson/node-jsonfile} \\
string-template     & 146   &	1.0.0   & {Matt-Esch/string-template} \\
wav-decoder         & 218   & 1.3.0   & {mohayonao/wav-decoder} \\ 
vtree               & 423   & 0.0.22  & {Matt-Esch/vtree} \\
underscore.string   & 873   & 3.3.4   & {esamattis/underscore.string}    \\
node-fs-extra       & 1274  & v11.1.1 & {jprichardson/node-fs-extra}    \\
messy               & 1694  & 6.11.2  & {papandreou/messy} \\
numbers.js          & 1780  &	0.6.0   & {numbers/numbers.js} \\
virtual-dom         & 1978  & v2.1.1  & {Matt-Esch/virtual-dom} \\
planck.js           & 10390 & v0.2.7  & {shakiba/planck.js} \\
ts-lib-mathjs       & 24645 & 3.0.0   & {em-tlm/ts-lib-mathjs} \\
goojs               & 57373 & v0.16.8 & {GooTechnologies/goojs} \\

\bottomrule
\end{tabular}

\captionsetup{justification=centering}
\caption{Open source projects used in the experimental study.}
\label{Tab:projectImplementationDetails}
\end{table}
\endgroup

Table~\ref{Tab:projectImplementationDetails} presents 
the projects used in our evaluation along with their implementation details. 
Column 2 (SLOC) provides the size of each project, in terms of line count (SLOC).
The SLOC metric is computed with the CLOC tool\footnote{https://github.com/kentcdodds/cloc}. 
Columns 3 and 4 provide the version of each project (Tag) 
and its Github repository name,
for study reproducibility reasons.

The proposed method is implemented as a Node.js prototype
and depends on static and dynamic analysis infrastructure.
Specifically, the prototype employs jscodeshift
\footnote{https://github.com/facebook/jscodeshift} 
for Abstract Syntax Tree (AST) traversal and code generation.
Moreover, it depends on Nodeprof
\footnote{https://github.com/Haiyang-Sun/nodeprof.js}, 
a Node.js framework for implementing dynamic analysis tools
in JavaScript~\cite{sun2018efficient}.
The implemented prototype requires limited user interaction, 
since the user only needs to specify the application and the system component, 
e.g. function, that is subject to test generation. 
The source code of the prototype implementations, 
as well as details on the reproduction of the results of the empirical study
are available as a Github repository\footnote{https://github.com/katerinapal/js-test-augmentation}.

\subsection{Evaluation Results}

\subsubsection{RQ1. What is the degree of test augmentation attained by the proposed method?}

The purpose of RQ1 is to investigate how effective is the proposed method
in augmenting a test suite with a sufficient number of unit tests,
given an initial seed of integration tests.
In order to answer RQ1,
at first, we identified through static analysis
all integration tests included in the test suites of benchmark projects.
These tests exercise components with at least one dependency
implemented in the project's codebase.
Then, we executed our prototype implementation
for each one of these components.
The augmented test suite for each project was committed
in a separate Git branch for further analysis.

Table~\ref{Tab:generatedTestsDetails} summarizes the results
generated by the analysis of the tests suites 
and the execution of our prototype.
Column 2 provides the total number of test cases
implemented in the test suites of the benchmark projects. 
Column 3 provides the number of test cases that serve as integration tests,
since they exercise components, not isolated from their dependencies.
Inside parentheses, we provide their ratio, 
over the total number of tests.
The number of generated unit tests, augmenting each test suite,
is presented in Column 4.
Finally, Column 5 provides the ``Augmentation Ratio'',
i.e., the percent increase in the size 
of the integration test suite (Column 3) 
due to the generated unit tests.

\begingroup
\setlength{\tabcolsep}{6pt} 
\renewcommand{\arraystretch}{1.2} 
\begin{table}[tb]
\caption{Test suite features and generated tests for benchmark projects.}
\label{Tab:generatedTestsDetails}
\centering
\scriptsize
\begin{tabular}{@{\extracolsep\fill}l r r r r@{}}
\toprule
\thead{Project} & 
\thead{Tests} &
\thead{Integration\\Tests (\%)} &
\thead{Generated\\Unit Tests} &
\thead{Augmentation\\ratio (\%)} \\ \hline
node-jsonfile     & 45    & \valpct{14}{31.11}  & 14  &  100.0 \\
string-template   & 87    & \valpct{38}{43.67}  & 88  &  231.57 \\
wav-decoder       & 3     & \valpct{1}{33.33}   & 1   &  100.0 \\ 
vtree             & 6     & \valpct{1}{16.66}   & 9   &  900.0 \\
underscore.string & 80    & \valpct{3}{3.75}    & 51  &  1700.0 \\
node-fs-extra     & 436   & \valpct{34}{7.79}   & 61  &  179.41 \\
messy             & 229   & \valpct{16}{6.98}   & 69  &  431.25 \\
numbers.js        & 140   & \valpct{14}{10.0}   & 56  &  400.0 \\
virtual-dom       & 121   & \valpct{24}{19.83}  & 43  &  179.16 \\
planck.js         & 14    & \valpct{2}{14.28}   & 4   &  200.0 \\
ts-lib-mathjs     & 3207  & \valpct{22}{0.68}   & 87  &  395.45 \\
goojs             & 1047  & \valpct{13}{1.24}   & 13  &  100.0 \\

\bottomrule
\end{tabular}

\end{table}
\endgroup

Based on the results of Table~\ref{Tab:generatedTestsDetails}, 
the share of integration tests in the test suites of benchmark projects
ranges from 0.68\% to 43.67\% with a median value of 12.14\%.
A high share of integration tests, e.g. in \texttt{string-template},
denotes that the test pyramid is, probably, inverted
and the test suite is deficient of unit tests.
In such cases, the proposed method can revert this situation
by augmenting the test suite with several additional unit tests.
Specifically, as indicated in Column 4,
test suites are augmented up to a maximum of 88 unit tests
with a median value of 47.
As concerning augmentation ratio,
it ranges from 100\% to 1700\% and a median value of 215\%.
An augmentation ratio 100\% represents the base case
where a single unit test is generated for each integration test,
e.g. in projects \texttt{node-jsonfile}, \texttt{wav-decoder}, \texttt{goojs}.
This indicates that exercised execution paths in integration tests
involve a single call site of dependent components.
On the other hand, a high augmentation ratio suggests 
the presence of multiple call sites of component dependencies 
in execution paths that seed the generation of unit tests.
For instance, an augmentation ratio of 1700\% in \texttt{underscore.string}
indicates that an average of 17 component dependencies are invoked 
in each one of its 3 integration tests.

\begin{boxwithhead}{Answer to RQ1}
{
The effectiveness of the proposed method 
depends, primarily, on the stratification of test types
in the test suite of a project.
A high number of integration or system tests
provides the potential for generating several unit tests.
This, also, depends on the number of call sites of component dependencies
covered by the execution of the integration tests.
The empirical evaluation of test generation effectiveness
reveals an augmentation ratio that ranges from 100\% to 1700\%,
with a median value of 215\%.
This suggest the potential of our method to generate 
a satisfactory number of unit tests, 
ranging from 1 to 17 unit tests for each integration test.
}
\end{boxwithhead}

\subsubsection{RQ2. Does the proposed method enhance the fault detection capability of a test suite?}

The purpose of RQ2 is to evaluate the practicality of the proposed method,
as concerning the capability of generated tests to uncover bugs
not detected by the original test suite.
Notice that the augmented test suite does not increase code coverage,
since generated tests exercise subpaths of execution paths
already covered by the integration tests.
However, each generated test introduces assertions
corresponding to intermediate program states,
enabling, thus, detection of faults whose effects
are not propagated to the result of its respective integration test.

In order to answer RQ2, 
we conduct an experimental study on the generated tests,
based on mutation analysis with the help of Stryker\footnote{https://stryker-mutator.io/}, 
a mutation analysis tool for JavaScript applications. 
The experimental setup involves 
execution of a two-stage mutation analysis.
In the first stage,
we perform mutation analysis on the integration tests of each codebase.
In a next stage, we repeat the mutation analysis
for the augmented test suites 
comprising the integration tests and the generated unit tests.
Notice that in the second stage,
the Stryker tool is appropriately configured 
in order to produce the same mutants during mutation analysis.

\begingroup
\setlength{\tabcolsep}{6pt} 
\renewcommand{\arraystretch}{1.2} 
\begin{table*}[tb]
  \centering 
\caption{Mutation analysis results in the selected benchmarks.}
\label{Tab:mutationAnalysisResults}
{
\scriptsize
\begin{tabular}{@{\extracolsep{2pt}}l r r r r r  r r r@{}}
\toprule
\multirow{2}{2.0cm}{\textbf{Project}} &
\multicolumn{3}{c}{\textbf{Integration Tests}} &
\multicolumn{4}{c}{\textbf{Augmented Integration Tests}} \\

\cmidrule{2-4} \cmidrule{5-8}
  & \thead{Killed} 
  & \thead{Survived}
  & \thead{Mutation\\Score (\%)}
  
  & \thead{Killed}
  & \thead{Survived}
  & \thead{Mutation\\Score (\%)}
  & \thead{Mutation Score\\Increase (\%)}\\ \hline

node-jsonfile     & 8     & 78    & 9.30      & 8     & 78    & 9.30  & 0.0 \\
string-template   & 179   & 35    & 83.64     & 179   & 35    & 83.64 & 0.0 \\
wav-decoder       & 105   & 78    & 57.38     & 106   & 72    & 59.55 & 3.79 \\ 
vtree             & 45    & 437   & 9.34      & 81    & 401   & 16.80 & 80.00 \\
underscore.string & 809   & 128   & 86.34     & 809   & 128   & 86.34 & 0.0 \\
node-fs-extra     & 96    & 9     & 91.43     & 96    & 9     & 91.43 & 0.0 \\
messy             & 1880  & 819   & 69.66     & 1880	& 819   & 69.66 & 0.0 \\
numbers.js        & 1683  & 470   & 78.17     & 1683  & 470   & 78.17 & 0.0 \\
virtual-dom       & 4	    & 7	    & 36.36     & 4     & 7     & 36.36 & 0.0 \\
planck.js         & 541   & 9780  & 5.24      & 593   & 9719  & 5.75  & 9.71 \\
ts-lib-mathjs     & 2722  & 26330 & 9.37      & 3329  & 25656 & 11.49 & 22.58 \\
goojs             & 865	  & 53    & 94.23     & 876   & 53    & 94.29 & 0.07 \\

\hline
\textbf{Total} & \textbf{8937} & \textbf{38226} & \textbf{52.54} & \textbf{9644} 
& \textbf{37447} & \textbf{53.57} & \textbf{9.68} \\

\bottomrule
\end{tabular}
}
\end{table*}
\endgroup

Table~\ref{Tab:mutationAnalysisResults} 
presents the results of the mutation analysis.  
Columns 2--4 provide the results of the first stage analysis.
Specifically, Columns 2 and 3 show
the number of killed and survived mutants, respectively, 
while Column 4 provides the mutation score.  
The second stage analysis results are available in Columns 5--7.
Columns 5 and 6 present 
the number of killed and survived mutants  
during the execution of the augmented test suite, 
and Column 7 provides the respective mutation score.
Finally, Column 8 illustrates the percent increase in the mutation score
for the augmented test suite.

Based on mutation analysis results, 
we observe that in 5 out of 12 projects
the generated unit tests kill more mutants 
and contribute to a percent increase to mutation score 
that ranges from 0.07\% to 80\%.
For instance in \texttt{vtree} project,
integration tests killed 45 mutants, 
while the augmented test suite killed 81,
resulting to 80\% increase of mutation score 
with respect to its initial value.
Moreover, in \texttt{ts-lib-mathjs} killed mutants
increase from 2722 to 3329 resulting 
to 22.58\% improvement to mutation score.

We attribute this project specific improvement to
(a) the type of returned results of exercised component dependencies
that mainly comprise object references and not primitive values,
(b) the capture and validation of internal program states by generated unit tests,
mainly in the form of assertions on changed attributes of returned objects.
Thus, changes to internal program state not propagated to the result
of integration tests contribute to increased mutation score.
On the other hand, in cases where mutation score is not affected
exercised component dependencies return primitive types and
their values probably have a direct contribution 
to the validated outcome of the target component.
For instance, in \stringtemplate{} project 
intermediate program states are dominated by string values
shaping the outcome of initially tested components.

\begin{boxwithhead}{Answer to RQ2}
{
The proposed method enhances, in several cases, 
the fault detection capability of a test suite,
despite maintaining the same code coverage.
This is due to the thorough validation of internal program states
enabled by the generated unit tests,
uncovering faults whose effects could not be captured 
by the integration tests.
The fault detection potential of the augmented test suite
is empirically evaluated through mutation analysis.
The results highlight an improvement to mutation score,
ranging from 0.07\% to 80\% in 5 out of 12 benchmark projects.
}
\end{boxwithhead}

\subsubsection{RQ3. Does the proposed method generate maintainable unit tests?}

In RQ3 our aim is to evaluate the extent 
to which the proposed method is capable of generating understandable 
and, thus, maintainable unit tests.
The focus is on the \emph{arrange} part of test cases and
the complexity of the code that prepares the test fixtures.

Our intuition is that the proposed method generates more understandable unit tests,
especially in cases that the test fixture setup involves
the construction of complex object structures.
Instead of resinstating the object state, captured during dynamic analysis,
with extensive assignment statements,
our method creates a backward slice of the act statement 
comprising statements from the integration test and the target component.
Thus, the developer's ability to reason on the intent of the test 
relies more on familiar code and less on multiple arbitrary values
assigned to object properties.
Figure~\ref{fig:ts-lib-mathjs-test-fixture} demonstrates
a generated unit test from project \tslibmathjs{}
and manually modified version of the same test 
with test fixture setup using property assignments.
The generated test fixture in Listing~\ref{lst:ts-lib-mathjs-generated-test} 
leverages the system's production and test code (lines 7--17),
while the modified test fixture in Listing~\ref{lst:ts-lib-mathjs-modified-test} 
is based on direct state initialization of the invocation arguments of \subsetF{} 
(lines 7--34).

\begin{figure}[htbp]
  \begin{minipage}[t]{.5\textwidth}
  
  \begin{lstlisting}[language=JavaScript,
      captionpos=t,caption={Generated unit test},
      basicstyle=\ttfamily\footnotesize,
      xleftmargin=1em, numbers = left, 
      label = lst:ts-lib-mathjs-generated-test]
test("_setsubset", () => {
  var math = require("../index");
  var _setsubset = 
  require("../subset_module.js")._setsubset;
  var index = math.index;
  var SparseMatrix = math.type.SparseMatrix;

  var m = new SparseMatrix([
          [0, 0],
          [0, 0]
        ]);
  var arg0 = m;
  var arg1 = index(1, 1);
  var arg2 = 1;
  var arg3 = undefined;
  var actualResult = 
    _setsubset(arg0, arg1, arg2, arg3);
  
  expect(actualResult._values["0"]).toBe(1);
  expect(actualResult._index["0"]).toBe(1);
  // ... more assertions
});
  \end{lstlisting} 

  \end{minipage}
    \begin{minipage}[t]{.5\textwidth}
    
      \begin{lstlisting}[language=JavaScript,
      captionpos=t,caption={Modified unit test},
      basicstyle=\ttfamily\footnotesize,
      xleftmargin=1em, numbers = left, 
      label = lst:ts-lib-mathjs-modified-test]
test("_setsubset", () => {
  var math = require("../index");
  var _setsubset = 
    require("../subset_module.js")._setsubset;
  var index = math.index;
  var SparseMatrix = math.type.SparseMatrix;

  var ImmutableDenseMatrix = 
    math.type.ImmutableDenseMatrix;
  var m = new SparseMatrix();
  m._values = [];
  m._index = [];
  m._ptr = [0, 0, 0];
  m._datatype = undefined;
  m._size = [2, 2];
  var arg0 = m;
  var arg1 = index(0,0);
  var arg1_dim0 = new ImmutableDenseMatrix();
  arg1_dim0._data = [1];
  arg1_dim0._size = [1];
  arg1_dim0._datatype = undefined;
  arg1_dim0._min = null;
  arg1_dim0._max = null;
  var arg1_dim1 = new ImmutableDenseMatrix();
  arg1_dim1._data = [1];
  arg1_dim1._size = [1];
  arg1_dim1._datatype = undefined;
  arg1_dim1._min = null;
  arg1_dim1._max = null;
  arg1._dimensions = [arg1_dim0, arg1_dim1];
  arg1._isScalar = true;
  var arg2 = 1;
  var arg3 = undefined;
  var actualResult = 
    _setsubset(arg0, arg1, arg2, arg3);
    
  expect(actualResult._values["0"]).toBe(1);
  expect(actualResult._index["0"]).toBe(1);
  // ... more assertions as above
});
  \end{lstlisting} 
  
  \end{minipage}
    \caption{Generated unit test and modified unit test with property assignments
    in the project \tslibmathjs{}.}
    \label{fig:ts-lib-mathjs-test-fixture}
\end{figure}

In order to answer RQ3 we conduct an experimental study
with purpose to compare unit tests generated by the proposed method 
against equivalent unit tests that use property assignments for test fixture setup.
The comparison is performed in terms of maintainability 
which is estimated with appropriate metrics. 
Since unit tests are not, normally, characterized by cyclomatic complexity,
we employed token-based metrics for estimation of maintainability
and, specifically, the complexity metrics proposed by Halstead~\cite{halstead1977elements}.
According to Halstead, 
software complexity can be estimated in terms of operators and operands,
where operators correspond to programming language tokens and names of user defined functions,
while operands are tokens introduced by the programmer (e.g. literals, variable names).
Among the metrics proposed by Halstead we opted to use \emph{Difficulty} (D)
that quantifies how difficult is to understand, write or maintain a code fragment.
As an additional proxy for maintainability, 
we used the \emph{Maintainability Index} metric (MI), 
used by Visual Studio to provide hints on code complexity. 
The MI metric computes maintainability as a function of 
Halstead Volume, Cyclomatic Complexity and Lines of Code
and is a simplified version of the metric proposed in~\cite{coleman1994using}.

The context of the experimental study for RQ3 comprises a subset of the benchmark projects
and, specifically, \texttt{vtree}, \texttt{planck.js}, \texttt{ts-lib-mathjs} and \texttt{goojs}. 
In these projects, setup of the test fixture in generated unit tests
involves appropriate initialization of object structures 
for the receiver object or the arguments of the act statement.
In the rest of benchmark projects, test fixture setup is simple, 
since the act statement receives primitive types as arguments.
For each generated unit test in the 4 benchmark projects,
we manually create an equivalent unit test 
where test fixture setup is based on object creation and property assignment statements.
We will, henceforth, refer to these tests as modified unit tests.
Especially for \texttt{ts-lib-mathjs}, characterized by several generated tests,
we randomly sampled 10\% of generated tests to include in this analysis.
Finally, we estimate Difficulty and MI metrics 
for both the generated and modified unit tests 
with the help of jsplato\footnote{https://github.com/caozhong1996/jsplato}, 
a static analysis tool for Halstead metrics estimation in JS source code. 

Table~\ref{Tab:qualityMetricsResults} presents the results of the experimental study. 
Columns 2 and 3 specify the cumulative values of Maintainability Index (MI) and Difficulty (D) metrics
for the generated unit tests,
while Columns 4 and 5 provide the respective values for the modified unit tests.
Moreover, Columns 6 and 7 present the percentage change in Maintainability Index and Difficulty
for modified unit tests with respect to generated tests.
Based on computed metric values,
we observe that the MI metric decreases across all projects for the modified unit tests,
denoting a corresponding decrease in maintainability.
As concerning Difficulty, it exhibits a significant increase in 3 out of 4 projects,
ranging from 14.63\% to 58.61\%.
Generally, the decrease in maintainability and understandability of modified tests
is due to the increased size of test fixtures in terms of lines of code
and total number of operators and operands.
Finally, the marginal decrease of \textit{Difficulty} 
in \texttt{goojs} benchmark is due to the fact that 
test fixtures in the generated tests specify array elements. 
Thus, these fixtures are based on the application's 
source code for defining the array instead of 
the value state of the array element 
which is leveraged in the modified tests.

\begingroup
\setlength{\tabcolsep}{6pt} 
\renewcommand{\arraystretch}{1.2} 
\begin{table}[htb]
  \centering 
\caption{Software complexity metrics for generated and modified unit tests.}
\label{Tab:qualityMetricsResults}
{
\scriptsize
\begin{tabular}{@{\extracolsep{0pt}}l r r r r r r@{}}
\toprule
\multirow{2}{2.0cm}{\thead{Project}} & 
\multicolumn{2}{c}{\thead{Generated\\Unit Tests}} &
\multicolumn{2}{c}{\thead{Modified\\Unit Tests}} &
\multicolumn{2}{c}{\thead{Percentage\\Change (\%)}} \\

\cmidrule{2-3} \cmidrule{4-5} \cmidrule{6-7} 
  & \thead{MI}
  & \thead{D}

  & \thead{MI}
  & \thead{D}

  & \thead{MI}
  & \thead{D}\\ \hline

vtree           & 61.75 & 51.23   & 59.79 & 61.15   & -3.17   & 19.36\\
planck.js       & 53.81 & 19.40   & 42.39 & 30.77   & -21.22  & 58.61\\
ts-lib-mathjs   & 42.98 & 103.39  & 39.26 & 118.52  & -8.66   & 14.63\\
goojs           & 53.69 & 28.31   & 50.31 & 26.48   & -6.30   & -6.46\\

\bottomrule
\end{tabular}
}

\end{table}
\endgroup

\begin{boxwithhead}{Answer to RQ3}
{
The proposed method generates unit tests 
with intent-revealing and intuitive code for test fixture setup, 
based on statements from the integration test 
and the target component implementation.
Generated unit tests are more maintainable 
than typical unit tests that initialize the state of required objects
with extensive property assignment statements.
This is also highlighted by the comparative improvement 
of \textit{Maintainability Index} and \textit{Difficulty} metrics
in generated unit tests with respect to manually created typical unit tests.}
\end{boxwithhead}


\section{Threats to validity} \label{threats-to-validity}

In this section, we describe potential threats 
to the validity of the empirical study evaluating the proposed method.
We prioritize validity threats based on the categorization of ~\cite{wohlin2012experimentation} as 
internal, external, construct and conclusion validity threats. \\

Threats to \emph{internal validity} involve factors 
affecting the credibility of cause-effect relationships 
supported by the results of the empirical stydy.
Since our study is exploratory and does not seek 
to confirm any causality relationships,
we deem that we can safely ignore such factors.

\emph{External validity} threats refer
to factors that may influence the generalization of the evaluation results. 
A probable threat is the extent at which 
the selected benchmark projects and their respective test suites
are representative of the population of Node.js applications. 
In order to mitigate this threat 
we carefully selected Github projects 
from different application domains 
characterized by a variety of sizes
in terms of production code SLOC 
and number of test cases in their test suites.

Threats to \emph{construct validity} regard issues influencing 
the relation between the studied theoretical constructs
and the observations of the empirical study. 
A potential threat involves the correct computation of the metrics 
involved in RQ2 and RQ3, to support the effectiveness and practicality 
of the proposed method. 
We handled this threat through adoption of widely used open-source tools
for the estimation of mutation score and maintainability metrics.
Stryker and jsplato are actively maintained by the developer community,
for several years, and thus we assume that they have reached
an adequate level of reliability in their results.

Threats to \emph{conclusion validity} are related to the statistical significance
of the observations and causality relationships
studied in the empirical study.
Since we conduct an exploratory study
that does not involve any hypothesis testing
we evaluate \emph{reliability} as a counterpart of conclusion validity~\cite{wohlin2012experimentation}.
A potential threat to reliability is a researcher bias 
on the extraction of data relevant to the application of the proposed method 
and their analysis to support study results.
We alleviate this threat through focusing on reproducibility of the results.
Towards this direction we selected publicly available open source projects as a context
and provide access to all required artifacts for study reproduction 
through a public Github repository.


\section{Related work} \label{relatedWork}


\subsection{Test refactoring} 
An empirical study on the impact of design 
patterns such as the Arrange-Act-Assert pattern 
test case readability is presented on~\cite{wei2025developers}. 
According to the authors, 
developers employ design patterns 
to design new test cases 
each of which focuses on verifying a single 
scenario of the application; 
however, they consider error-proneness 
before refactoring existing test cases. 
A systematic literature review on 
the quality of test assertions 
generated by use-of-the-art prototypes 
is proposed in~\cite{taromirad2025assertions}. 
Among the problems regarding test assertion generation, 
the lack of specification of the analyzed system's 
intended behaviour and error-proneness of the 
generated assertion oracles are noted. 

A number of techniques are proposed 
towards test refactoring for Java and Python projects. 
An approach that employs static and dynamic analysis 
techniques for refactoring rotten tests 
towards improving software quality 
is proposed in~\cite{martinez2020rtj}. 
A method for the extraction of test cases 
that improves dynamic analysis techniques such as
fault localization and software repair
is mentioned in~\cite{xuan2016b}. 
Initial system tests are split in 
smaller test cases which cover 
a more specific part of the control flow.  
Thus, new bugs are uncovered  
through examining traces 
which where hidden in the initial tests. 
The executed control flow is selected based
on the dynamic analysis technique specified 
for test extraction. 
A test minimization method  
which handles redundant test statements 
is proposed in~\cite{vahabzadeh2018fine}. 
The proposed method groups 
the initial system tests based on their fixtures; 
each test group is mapped to a new test 
examining the program traces 
uncovered by the test group, 
thus improving regression testing 
through reducing test suite execution time. 
Contrary to these methods, 
our method generates new assertions 
based on program state, 
thus improving fault localization 
through uncovering execution traces that 
were not examined in the initial test assertions. 
A technique for refactoring Python test clones, 
namely tests
that verify a single scenario with different inputs, 
towards generating parameterized tests is proposed in~\cite{kingston2024automated}. 
The authors state that the proposed method is 
able to produce effective tests except for 
tests containing invocations to other tests. 

\subsection{Regression testing}
A number of studies have been recommended towards 
regression testing for Java and Python programs~\cite{robinson2011scaling, lukasczyk2022pynguin, lukasczyk2023empirical}. 
In~\cite{robinson2011scaling},  
feedback-directed random testing techniques are leveraged  
in order to iteratively construct unit tests 
whose execution traces guide test generation. 
In~\cite{lukasczyk2022pynguin, lukasczyk2023empirical}, 
search-based testing techniques, 
such as inferring type declarations through inspecting 
the context in which a program variable is referenced, 
are employed in order to uncover the test inputs which 
expose program behavior. 
In terms of inferring the analyzed program behavior, 
our method relies on existing test cases for 
generating more specific tests 
instead of iteratively generating unit tests.

\subsection{Large Language models (LLMs)} 
A study evaluating the integration of static analysis 
techniques in LLM-based test generation methods is 
presented in~\cite{pan2025aster}. 
The authors state that static analysis techniques 
improve the tests provided by LLMs in terms of 
coverage and understandability. 
Several studies focus on employing 
natural language processing methods in 
unit test generation for Java, C++ and 
Python~\cite{10.1145/3660783, yang2024empirical, bhatia2024unit, zhang2025citywalk}. 
A method for generating unit tests for Java projects 
is proposed in~\cite{10.1145/3660783}. 
The method infers the intention of the focal method 
which is then used to guide test generation 
instead of the method's source code. 
Finally, the method leverages the generated 
tests' compilation error messages to improve test correctness. 
A study on the effectiveness of advanced language models 
in unit test generation for Java projects 
is presented in~\cite{yang2024empirical}. 
The authors conclude that LLM effectiveness 
relies on prompting, 
which involves describing code features 
in order to facilitate the generation 
of syntactically correct tests. 
A technique that leverages both 
Large Language Models and program analysis 
to generate tests for C++ programs is 
proposed in~\cite{zhang2025citywalk}. 
Specifically, program analysis techniques 
are applied to capture project-specific information 
such as component dependencies, 
leading to more effective and accurate test cases. 
A comparison between Large Language Models and 
state-of-the-art unit test generators for 
Python programs is presented in~\cite{bhatia2024unit}. 
The authors show that LLMs outperform unit test generators 
in terms of performance.    
However they produce incorrect assertions, 
since they are based on natural language processing 
methods instead of code semantics.  
An iterative LLM-based test generation technique for 
Python systems is presented in~\cite{jain2025testforge}. 
The technique generates a test suite for a source code file, 
which is then refined to target code segments with small mutation scores
based on execution feedback, such as runtime failures. 
It outperforms state-of-the-art test generation methods 
in terms of providing more readable tests 
since it combines LLM-based techniques with feedback 
from the execution of the generated tests. 
A LLM-based technique that leverages callgraphs to 
generate tests for Python programs is proposed in~\cite{liu2025llm}. 
The method leverages the invocations of a function under test 
to infer the function parameter types, 
thus providing more effective tests. 

\subsection{Fault localization}

Several techniques were proposed towards improving fault localization, 
in terms of dividing test cases into smaller fractions and providing test oracles. 
A test case purification technique for locating software faults in Java programs
which is based on the system's failing tests is proposed in~\cite{xuan2014test}. 
The method integrates each assertion omitted 
after the failure of a unit test in a new test, 
thus enabling further fault localization. 
A technique that combines program analysis and 
LLMs towards isolating faults in C programs is 
presented in~\cite{farzandway2025automated}. 
The proposed technique leverages an iterative 
prompting approach based on code patch generation, 
reasoning on the execution result of the generated 
patches and using reasoning outcomes in 
the next iterations. 
The authors state that the technique outperforms 
state-of-the-art fault localization techniques 
in terms of repair accuracy. 
An empirical study for comparing the effectiveness of 
dynamic slicing techniques against 
statistical fault localization in C programs 
is presented in~\cite{soremekun2021locating}. 
The authors conclude that statistical formulas 
(1) are effective in locating faults by considering only the 
most suspicious source location, 
(2) outperform dynamic slicing techniques, in terms of 
locating faults in a larger set of benchamrks 
through providing a small set (5\%) of the suspicious source locations,
(3) are more time-consuming, in terms of analyzing 
more code than dynamic slicing 
which considers only faulty executions. 
Thus, a hybrid method, i.e. 
inspecting dynamic slices starting from 
suspicious statements reported by statistical formulas, 
is more effective in terms of effectiveness and time. 

\subsection{Test amplification}
A test amplification technique for Pharo programs 
is presented in~\cite{abdi2022small}. 
The technique employs test execution traces 
in order to discover the program's type declarations, 
which are useful for generating test cases 
that uncover edge execution cases 
of the program under test. 
However, our method involves generating 
test cases that enhance reasoning on 
existing software faults, 
instead of uncovering faulty behavior 
based on the behavior verified by 
the program's test cases. 

Another technique for test amplification on Python programs 
is proposed in~\cite{schoofs2022ampyfier}. 
The technique generates new test assertions 
in order to uncover bugs in program traces 
not examined in the initial test, 
through tracking object values during program execution. 
Compared to this method, 
our method involves generating 
test fixtures using the system's initial 
production and test code through leveraging 
backward program slicing, 
as described in Section~\ref{method}, 
thus improving software readability and understandability.

An integration of a test amplification tool 
into the tools used in software development tasks 
is proposed in~\cite{abdi2024test}. 
The authors state that integrating 
amplification tools in common development tasks 
should involve test prioritization 
in order to provide insights about test code quality 
in a reasonable amount of time. 
A LLM-based technique for amplifying 
existing test suites of REST APIs is 
proposed in~\cite{nooyens2025test}. 
According to the authors, the proposed method 
outperforms state-of-the-art testing methods 
focusing on REST APIs through 
iteratively refining the generated tests based 
on execution feedback such as runtime failures. 

\subsection{Test transplantation}
A technique for generating unit tests for Pharo libraries 
based on the projects that depend on these libraries 
is presented in~\cite{abdi2022test}. 
Dynamic slicing is leveraged to generate 
test fixtures from the dependent project test suites. 
The generated fixtures are then integrated 
in the tests examining library methods. 
A method for generating tests with mock objects for 
Java projects is presented in~\cite{tiwari2024mimicking}. 
The method leverages static analysis methods 
to identify the methods invoked from the method under test; 
the identified method calls are then instrumented 
to capture information about their parameters and results. 
The authors show that the generated tests 
can adequately mimic the observed behavior of the method 
under test since they can recreate its invocation context 
through the generated test fixtures and oracles. 


\section{Conclusions and Future work} \label{conclusion}

We have proposed a method that employs static and dynamic analysis
for augmenting a test suite with automatically generated unit tests.
The method is most suitable for test suites
where the stratification of unit, integration and system tests 
does not conform to the recommended test pyramid structure:
numerous unit tests providing high code coverage and forming the base,
fewer integration tests in the middle that verify component collaboration,
and far fewer system or UI tests at the top 
that exercise acceptance or other scenarios of use.
Instead, integration and system tests represent the majority of test cases,
resulting in coarse-grained tests 
with limited fault localization capability and longer execution times.
The method leverages integration tests, 
exercising a component and its dependencies,
to generate unit tests that verify component dependencies in isolation.
Specifically, given a target component, i.e., a function or method,
the proposed method generates one unit test 
for each call site and execution context of each one of its dependencies.
The call site of a dependency represents the \emph{Act} part of the test case,
while the code required for test fixture setup (the \emph{Arrange} part) 
is derived from a backward slice combined with program state captured through dynamic analysis.
The generation of the \emph{Assert} part is based on the invocation result
in the given call site, which is, also, tracked through dynamic analysis.

We showcase and empirically evaluate the proposed method in the Node.js platform, 
although it can be ported and adapted to other languages and platforms.
The evaluation is based on a research prototype implemented as a Node.js tool
and is conducted in the context of twelve open source JS applications (benchmark projects).
Evaluation results support the effectiveness and practicality of our approach.
Although the effectiveness of test generation 
depends on the prevalence of integration tests in the benchmark projects,
the proposed method achieved an augmentation ratio ranging from 100\% to 1700\%,
i.e., it generated 1 to 17 unit tests per integration test.
Moreover, mutation analysis on the integration tests 
and the augmented integration tests of each benchmark project
revealed an increase in fault detection capability due to generated unit tests.
Specifically, mutation score improved from 0.07\% to 80\% in 5 out of 12 benchmark projects,
which underpins the practicality of the approach.
The practicality is, also, enhanced due to the form of test fixture setup code,
especially, in cases that it involves initialization of complex object structures.
The proposed method employs statements from the target component and the integration test
for test fixture setup, resulting to intent-revealing and intuitive code.
On the other hand, typical unit tests initialize the required state 
through extensive property assignments.
This is highlighted by the comparative improvement of \emph{Maintainability Index} 
and \emph{Difficulty} metrics in generated unit tests with respect to manually created typical unit tests.

Future extensions of this work would involve porting the method to other platforms and languages, 
e.g. the Java platform. Moreover, we consider extensions that would enable unit test generation
during the actual usage of the system under test, 
resulting to test cases that capture realistic scenarios and test data.

\bibliography{bibliography}

\end{document}